\documentclass[aps,prd,superscriptaddress,onecolumn]{revtex4}
\usepackage{amssymb}
\usepackage{amsmath}
\usepackage{amsfonts}
\usepackage{epsfig}
\usepackage{graphicx}
\usepackage{tabularx}
\usepackage{color}
\usepackage{dcolumn}

\newcommand{\beaa}{\begin{eqnarray*}}
\newcommand{\enaa}{\end{eqnarray*}}
\newcommand{\bea}{\begin{eqnarray}}
\newcommand{\ena}{\end{eqnarray}}
\newcommand{\seq}{\begin{subequations}}
\newcommand{\sen}{\end{subequations}}
\newcommand{\eq}{\begin{eqnarray}}
\newcommand{\en}{\end{eqnarray}}

\def\shiftdown#1{#1\llap{\lower.04ex\hbox{#1}}}

\def\nn{\nonumber}

\def \be  {\begin{equation}}
\def \ee  {\end{equation}}
\def \ba  {\begin{eqnarray}}
\def \ea  {\end{eqnarray}}
\def \baa {\begin{eqnarray*}}
	\def \eaa {\end{eqnarray*}}

\def \bb  {\begin{thebibliography}}
	\def \eb  {\end{thebibliography}}

\def \lab #1 {\label{#1}}

\def \qqqquad {\qquad\qquad}

\def \e {\mbox{e}}

\begin{document}
	
\title{Transverse momentum dependence of the $T$-even hadronic \\
  structure functions in the Drell-Yan process}

\author{Valery E. Lyubovitskij} 
\affiliation{Institut f\"ur Theoretische Physik, 
Universit\"at T\"ubingen,  
Auf der Morgenstelle 14, D-72076 T\"ubingen, Germany}
\affiliation{Millennium Institute for Subatomic Physics at  
the High-Energy Frontier (SAPHIR) of ANID, \\  
Fern\'andez Concha 700, Santiago, Chile}  
\author{Alexey S. Zhevlakov}  
\affiliation{Bogoliubov Laboratory of Theoretical Physics,
JINR, 141980 Dubna, Russia} 
\affiliation{Matrosov Institute for System Dynamics and 
Control Theory SB RAS, \\
Lermontov str., 134, 664033, Irkutsk, Russia}
\author{Iurii A. Anikin}
\affiliation{Department of Physics, Tomsk State University,
634050 Tomsk, Russia}

	\date{\today}

\begin{abstract}
          
We present detailed analysis of the $T$-even lepton angular distribution
in the Drell-Yan process including $\gamma/Z^0$ gauge boson exchange
and using perturbative QCD based on the collinear factorization scheme
at leading order in the $\alpha_s$ expansion. We focus on the study of
the transverse momentum $Q_T$ dependence of the corresponding hadronic
structure functions and angular coefficients up to next-to-next-to-leading
order in the $Q_T^2/Q^2$ expansion. We analyze $Q_T$ dependence numerically
and compare $T$-even angular coefficients integrated over rapidity with
available data of the ATLAS Collaboration at LHC. Additionally, we present
our results for the forward-backward asymmetry and compare it with data.

\end{abstract}	

\maketitle
	
\section{Introduction}	

Study of hadron structure is one of the most attractive topics during
the last decades. In this vein, the Drell-Yan (DY) process~\cite{Drell:1970wh} 
$H_1(P_1) H_2(P_2) \to \ell_1 \bar{\ell}_2 X$ is one of the key tools
for getting a new information on hadronic structure functions.
In particular, theoretical analysis of data on angular distributions
of leptonic pairs ($\ell_1 \bar{\ell}_2$) gives direct access to
these physical quantities. 
During last ten years, angular distributions were measured at LHC
by the ATLAS~\cite{ATLAS:2016rnf}, CMS~\cite{CMS:2015cyj},
and LHCb~\cite{LHCb:2022tbc} Collaborations in large interval of
transverse momentum $Q_T$ of gauge boson producing leptonic pair.
Before the LHC era, measurements of angular distributions in the DY processes
have been done by the NA10~\cite{NA10:1986fgk}, NA3~\cite{NA3:1981flw},
and CDF~\cite{CDF:2011ksg} Collaborations. Advantage of new measurements
done by the ATLAS~\cite{ATLAS:2016rnf} and CMS~\cite{CMS:2015cyj}
Collaborations consists an extension of the study of DY processes
to the weak sector,   
an extension of the DY processes to the weak sector,
i.e. to study weak boson production.
Latest advanced calculations of the DY angular distributions/coefficients
give important opportunity for high-precision test of electroweak sector
of the Standard Model (SM). In particular, a precision of
the inclusive and full differential DY cross sections have been
extended from  next-to-next-to leading order (NNLO)~\cite{Hamberg:1990np,vanNeerven:1991gh,%
Anastasiou:2003yy,Melnikov:2006di,Catani:2009sm,Gavin:2010az,Gavin:2012sy} 
to N$^3$LO~\cite{Duhr:2020seh,Baglio:2022wzu,Camarda:2021jsw}.
Besides, part of calculation include also parton
showers~\cite{Hoche:2014uhw,Karlberg:2014qua,Alioli:2015toa}.
Also one should mention that the success of the 
parton Reggeization approach~\cite{Nefedov:2012cq}-\cite{Nefedov:2020ugj}
for study DY processes at high energies.
Careful and consistent inclusion of the electroweak corrections have been
made in Refs.~\cite{Baur:2001ze,%
Dittmaier:2001ay,Arbuzov:2005dd,Arbuzov:2007db,CarloniCalame:2007cd}. 
For a status of the  QCD precision predictions
for the DY processes see, e.g., ~\cite{Alekhin:2024mrq}. 

Current measurements allow for the accurate verification of theoretical
predictions regarding the behavior of angular distributions
at substantial transverse momentum $Q_T$. 
In perturbative QCD (pQCD) one can systematically
predict $Q_T$ dependence of the structure functions in order-by-order
in the strong coupling $\alpha_s$ expansion.  
Analysis of the $Q_T$ dependence of the angular distributions in the
electromagnetic DY process at order $\alpha_s$ in the collinear factorization
scheme was made in Refs.~\cite{Lam:1978pu,Lam:1978zr,%
Collins:1978yt,Boer:2006eq,Berger:2007jw}. For analysis of the angular
distributions of the DY leptons in the TMD factorization approach see
recent paper~\cite{Piloneta:2024aac}.  
Studies of angular coefficients with comparison  
with existing data were presented~\cite{Frederix:2020nyw,Motyka:2016lta,%
Peng:2014hta,Ebert:2020dfc,Gauld:2017tww,Mirkes:1992hu}.
Recent data of the ATLAS~\cite{ATLAS:2016rnf} and CMS~\cite{CMS:2015cyj}
Collaborations were analyzed, in particular, at the $O(\alpha_s^2)$ order
in strong coupling expansion by using
DYNNLO~\cite{Catani:2009sm} and FEWZ~\cite{Gavin:2010az} generators.
These packages retain full kinematical information about the final state
and allow for a direct comparison to data in the fiducial region.
In particular, using these generators, and later on the
NNLOJET~\cite{Gauld:2017tww} package, angular coefficients were extracted
by using methods proposed in Ref.~\cite{Mirkes:1994eb}, based on orthogonality
of harmonic polynomials and on connection to angular distributions.
The method of Ref.~\cite{Mirkes:1994eb} was based on integration over
the full phase space of the angular distributions. It cannot be applied
directly to data, but it was used to compute all the theoretical predictions,
in particular, based on Monte Carlo generators.  

In this paper we analyze DY angular distributions based on
collinear pQCD~\cite{Collins:1984kg,Collins:2011zzd}.
In particular, we focus on the small $Q_T$ limit,
which is important due to several reasons:
(1) for deeper understanding of this limit in the collinear pQCD; 
(2) for performing resummation of the hadronic structure functions at small
$Q_T$ proposed in Ref.~\cite{Berger:2007jw}. 
We derive analytical results for the helicity hadronic structure functions
up to next-next-to-leading power in the $Q_T^2/Q^2$ expansion. 
We also perform a comparison of our predictions for
the angular coefficients with the ATLAS data~\cite{ATLAS:2016rnf}
in the fiducial region with lepton pair produced in vicinity of
the $Z$-boson mass. Note, analysis of the $Q_T$ dependence of
the angular distributions was proposed and developed before
in Refs.~\cite{Collins:1984kg,Mirkes:1992hu,Berger:2007jw,Boer:2006eq}
in the leading order in the $Q_T^2/Q^2$ expansion and
recently in Ref.~\cite{Lyubovitskij:2024civ}
up to next-next-to-leading order in the $Q_T^2/Q^2$ expansion
for the case of the $T$-odd angular distributions in the DY process.
Comprehensive discussion of the formalism of the $Q_T^2/Q^2$ expansion
up to arbitrary order of accuracy can be found in Ref.~\cite{LVWZ:2023}.
This is good starting point for our study in the present paper, where we
focus on the: 
(1) performing the small $Q_T$ expansion of the $T$-even hadronic
structure functions up to next-next-to-leading order; 
(2) making numerical analysis of angular coefficients with data.
In our consideration we will deal with the DY cross processes
involving both photon and $Z^0$-boson productions.
Our numerical analysis includes a possible uncertainties of initial conditions,
such as the ranges of measured $Q$ or $Q_T$ for invariant mass of lepton pair. 

Our paper is organized as follows.
In Sec.~\ref{Helicity} we present the definition of the hadronic structure functions
and kinematics defining helicity structure functions. In Sec.~\ref{pert_calc} we present
our results for all $T$-even structure functions in the  $\alpha_s$ order in the framework of
collinear pQCD. In Sec.~\ref{small_Qt} we discuss the $Q_T^2/Q^2$ expansion 
up to next-next-to-leading order. In Sec.~\ref{Discussion} we present our numerical results
for the $Q_T$ dependence of the hadronic structure functions and compare them with
ATLAS data. We also discuss our predictions for the forward-backward (FB) asymmetry and
for the convexity (transverse-longitudinal hadronic structure asymmetry) and compare
it with available data. 
Finally, in Sec.~\ref{Conclusion} we give our conclusion.
In the Appendixes we collect some calculation details. In particular,
in Appendix~\ref{app:Kinematics} we discuss details of kinematic of the DY process.
In Appendix~\ref{app:Wstr} we include details regarding hadronic and leptonic
helicity structure functions. In Appendix~\ref{app:Str_Func} 
we show relations between three different sets of hadronic structure functions. 
In Appendix~\ref{app:Expansion}, we present perturbative coefficients
parametrizing small $Q_T^2/Q^2$ expansion of the hadronic structure functions. 
 
\section{Hadronic structure functions in the Drell-Yan process} 
\label{Helicity}

The DY process is specified as 
$H_1(P_1) + H_2(P_2) \to \gamma^*(Z^0) +X \to \ell^-(q_1) + \ell^+(q_2) + X$,
where $H_1$ and $H_2$ are the initial-state hadrons, 
$(\ell^+ \ell^-)$ is leptonic pair, $q = q_1 + q_2$ is the 
vector boson momentum; $d\Omega = d\!\cos\theta \, d\phi$ is the solid angle 
of the lepton $\ell^-(q_1)$ in terms of its polar $(\theta)$ and azimuthal 
$(\phi)$ angles in the center-of-mass (c.m.) system of the leptonic 
pair. The details of the kinematics of the DY process 
are given in Appendix~\ref{app:Kinematics}.
Leptonic c.m. frame is defined as 
\eq 
q &=& q_1 + q_2 = Q \, (1, 0, 0, 0)\,, \nonumber\\
k &=& q_1 - q_2 = Q \, (0, \cos\phi \sin\theta, \sin\phi \sin\theta, \cos\theta) \,. 
\en
The starting point for the study of the DY reaction is the
differential cross section defined in the form of contraction of 
lepton $L_{\mu\nu}$ and hadronic $W^{\mu\nu}$ tensors, 
\eq 
\frac{d\sigma}{d\Omega \, d^4q} = \frac{\alpha^2}{2 (2\pi)^4 Q^4 s^2} \,  
L_{\mu\nu} \, W^{\mu\nu} \,,
\en 
where 
$s=(P_1 + P_2)^2$ is the hadron-level total energy, 
$\alpha = 1/137.036$ is the electromagnetic fine structure constant, 
$Q^2 = q^2$ is the timelike vector boson momentum squared. 

In the expansion of the hadronic tensor $W^{\mu\nu}$ it is convenient to
use the helicity formalism proposed
in Ref.~\cite{Lam:1978pu} for reactions with photon exchange and
extended in Ref.~\cite{Mirkes:1992hu} to the electroweak case. 
In Ref.~\cite{Lyubovitskij:2024civ} we showed that the results
of Ref.~\cite{Mirkes:1992hu} for the expansion of $W^{\mu\nu}$
can be conveniently rewritten using a basis of orthogonal unit vectors
$T^\mu = q^\mu/\sqrt{Q^2} = (1,0,0,0)$, 
$X^\mu = (0,1,0,0)$, 
$Z^\mu = (0,0,0,1)$, 
$Y^\mu = \epsilon^{\mu\nu\alpha\beta}
T_\nu Z_\alpha X_\beta =(0,0,1,0)$, proposed in Ref.~\cite{Lam:1978pu}
and related to the hadron and virtual-boson momenta:
\eq 
P_1^\mu &=& e^{-y} \, \frac{\sqrt{s}}{2} \, 
\left( 
T^\mu \sqrt{1+\rho^2} + Z^\mu - \rho X^\mu \right) \,, \nonumber\\
& &\\
P_2^\mu &=& e^{y} \, \frac{\sqrt{s}}{2} \, 
\left( 
T^\mu \sqrt{1+\rho^2} - Z^\mu - \rho X^\mu \right) \,, \nonumber
\en
and polarization vectors for both photon and weak bosons ($G=W^\pm, Z^0$) are
\eq 
\epsilon^\mu_\pm(q) &=& \frac{\mp X^\mu - iY^\mu}{\sqrt{2}} \,, 
\nonumber\\[2mm] 
\epsilon^\mu_0(q) &=&Z^\mu \,.
\en
Here the hadronic momenta are chosen in the Collins-Soper frame
and related to the parton momenta 
$p_i = \xi_i P_i$, where $\xi_i$ is the partonic momentum
fraction,  $Q^+$, $Q^-$, $Q_T$ are the gauge boson
longitudinal and transverse momentum components, respectively,
with $Q^\pm = x_{1,2} \sqrt{s/2} = e^{\pm y} \sqrt{(Q^2+Q_T^2)/2}$. 
We introduce the following notations:
$\rho = Q_T/Q$ is the ratio of the transverse component and magnitude
$Q=\sqrt{Q^2}$ of the vector boson momentum, 
$x_{1,2} = 2 P_{1,2}q/s$ are the momentum fractions of the light cone
components of the finale vector boson, 
$y = (1/2) \log(x_1/x_2)$ is the rapidity. 
We also define the $x$ fraction factors at $Q_T^2 = 0$ as
$x_{1,2}^0 = e^{\pm y} (Q/\sqrt{s})$.  Tensor 
$\epsilon^{\mu\nu\alpha\beta}$ is the four-dimensional
Levi-Civita tensor defined via ${\rm tr}(\gamma^5 \gamma^\mu \gamma^\nu\gamma^\alpha
\gamma^\beta) = 4 \, i \, \epsilon^{\mu\nu\alpha\beta}$,
with $\epsilon^{0123} = - \epsilon_{0123} = -1$.  

The gauge boson vectors satisfy to the Lorentz condition, orthonormality 
and completeness conditions: 
\eq 
& &q_\mu \, \epsilon^\mu_\lambda(q) = 0\,, \quad 
\epsilon_{\mu,\lambda}(q) \, \epsilon^\mu_{\lambda^\prime}(q) 
= - \delta_{\lambda\lambda^\prime} \,, \nonumber\\[2mm] 
& &
\sum\limits_{\lambda = 0, \pm} 
\epsilon^\mu_\lambda(q) \, \epsilon^{\ast\nu}_\lambda(q) = 
- g^{\mu\nu} + \frac{q^\mu q^\nu}{q^2}  = 
- g^{\mu\nu} + T^\mu T^\nu = X^\mu X^\nu + Y^\mu Y^\nu + Z^\mu Z^\nu \,  .
\label{comp_cond} 
\en

The expansion of the hadronic tensor the basis of unit vectors
$X,Y,Z$ reads~\cite{Lyubovitskij:2024civ}
\eq\label{XYZ_helicity}
W^{\mu\nu} 
&=& 
(X^\mu X^\nu + Y^\mu Y^\nu) W_T 
\,+\, i (X^\mu Y^\nu - Y^\mu X^\nu) W_{T_P} 
\,+\,    Z^\mu Z^\nu W_L \nonumber\\[2mm]
&+&    (Y^\mu Y^\nu - X^\mu X^\nu) W_{\Delta\Delta}  
\,-\,   (X^\mu Y^\nu + Y^\mu X^\nu) W_{\Delta\Delta_P}\nonumber\\[2mm]  
&-&    (X^\mu Z^\nu + Z^\mu X^\nu) W_{\Delta} 
\,-\,   (Y^\mu Z^\nu + Z^\mu Y^\nu) W_{\Delta_P}\nonumber\\[2mm]   
&+&  i (Z^\mu X^\nu - X^\mu Z^\nu) W_{\nabla} 
\,+\, i (Y^\mu Z^\nu - Z^\mu Y^\nu) W_{\nabla_P}
\,. 
\en 
Here five structure functions $W_i$ ($i=T$, $L$, $\Delta\Delta$, 
$\Delta$,$\nabla$) are generated by parity-even part 
of the hadronic tensor $W_{\mu\nu}$, while the other four ones $W_i$ 
($i=T_P$, $\Delta\Delta_P$, $\Delta_P$, $\nabla_P$)  
by the parity-odd part of $W_{\mu\nu}$. They are classified as: 
two transverse functions  --- 
parity-even $W_T$ and parity-odd $W_{T_P}$, 
one longitudinal function --- $W_L$ (it is parity-even), 
two transverse-interference (double-spin-flip) functions --- 
parity-even $W_{\Delta\Delta}$ and 
parity-odd $W_{\Delta\Delta_P}$, 
four transverse-longitudinal-interference (single-spin-flip) 
functions --- parity-even $W_{\Delta}$, $W_{\nabla}$ and 
parity-odd $W_{\Delta_P}$, $W_{\nabla_P}$. 

The lepton angular distribution, which encodes the information about 
the polar and azimuthal asymmetries can be expanded in terms of the  
nine helicity structure functions $W_i$ corresponding to the specific
polarization of gauge boson~\cite{Lam:1978pu,Collins:1977iv,Mirkes:1992hu,%
Boer:2006eq,Berger:2007jw,Lyubovitskij:2024civ} (see details
in Appendix~\ref{app:Wstr})
\eq\label{dNdOmega} 
\frac{dN}{d\Omega} &=& \frac{d\sigma}{d\Omega d^4q} \, 
\biggl(\frac{d\sigma}{d^4q}\biggr)^{-1}
\nonumber\\
&=& \frac{3}{8 \pi (2 W_T + W_L)} \,
\biggl[ 
      g_T             \, W_T
\,+\, g_L             \, W_L 
\,+\, g_\Delta         \, W_\Delta
\,+\, g_{\Delta\Delta}  \, W_{\Delta\Delta}
\nonumber\\
&+&   g_{T_P}          \, W_{T_P}
\,+\, g_{\nabla_P}      \, W_{\nabla_P}
\,+\, g_{\nabla}        \, W_{\nabla}
\,+\, g_{\Delta\Delta_P} \, W_{\Delta\Delta_P} \,
\,+\, g_{\Delta_P}      \, W_{\Delta_P} \biggr] \,,
\en
where
$g_i = g_i(\theta,\phi)$ are the angular coefficients 
\eq\label{gcoeffs}
g_{_T} & = &  1 + \cos^2\theta\,,  \quad \hspace*{.6cm}
g_{_L} \, = \, 1 - \cos^2\theta\,,  \quad \hspace*{.45cm}
g_{_{T_P}} \, = \, \cos\theta\,, \nonumber\\
g_{_{\Delta\Delta}} & = & \sin^2\theta \, \cos 2\phi\,, \quad \hspace*{.2cm}
g_{_{\Delta}} \, =  \, \sin 2\theta \, \cos\phi\,, \quad  \hspace*{.15cm}
g_{_{\nabla_P}} \, =  \, \sin\theta \, \cos\phi\,,\nonumber \\ 
g_{_{\Delta\Delta_P}} & = & \sin^2\theta \, \sin 2\phi\,, 
\quad  \hspace*{.04cm}
g_{_{\Delta_P}} \, = \, \sin 2\theta \, \sin\phi\,, \quad 
\hspace*{.4cm}
g_{_{\nabla}} \, = \,\sin\theta \, \sin\phi \,.
\en
Note, the six angular coefficients $g_i$ 
($i = T$, $L$, $\Delta\Delta$, $\Delta$, $\Delta\Delta_P$, $\Delta_P$) 
are invariant under $P$-parity transformation 
$\theta \to \pi - \theta$ and $\phi \to \pi + \phi$, while 
the other three coefficients $g_{i}$ ($i=T_P$, $\nabla$, 
$\nabla_P$) change the sign in that case~\cite{Mirkes:1992hu}. 
Hence, the six partial lepton angular distributions $dN_i/d\Omega$ \ 
($i = T$, $L$, $\Delta\Delta$, $\Delta$, $T_P$, $\nabla_P$) 
are the $P$-parity invariant, 
while the other three distributions $dN_i/d\Omega$ 
($i = \Delta\Delta_P$, $\Delta_P$, $\nabla$) are the $P$- and
also $T$- parity odd, which are generated at next-to-leading order
by the absorptive part 
of the parton scattering amplitude~\cite{Mirkes:1992hu,Lyubovitskij:2024civ}.
Recently in Ref.~\cite{Lyubovitskij:2024civ} we studied in detail
$T$-odd angular distributions in the case of the DY reactions.
In present paper we focus on the $T$-even angular distributions. 

There are two other commonly employed, and equivalent,
parametrizations of the lepton angular distribution in
literature~\cite{Lam:1978pu,Collins:1977iv,Mirkes:1992hu,%
Boer:2006eq,Berger:2007jw}
\eq \label{eq4-sf}
\frac{dN}{d\Omega} &=& \frac{3}{16\pi} \, 
\biggl( 
1 + \cos^2\theta 
+ \frac{A_0}{2} (1-3\cos^2\theta) 
+ A_1 \sin 2\theta  \cos\phi 
+ \frac{A_2}{2} \sin^2\theta  \cos 2\phi \nonumber\\
&+& A_3 \sin\theta  \cos\phi 
+ A_4 \cos\theta  
+ A_5 \sin^2\theta \sin 2\phi 
+ A_6 \sin 2\theta \sin\phi 
+ A_7 \sin\theta  \sin\phi  
\biggr)\,,
\en
and 
\eq\label{dNGreek}
\frac{dN}{d\Omega} &=& \frac{3}{4\pi} \, \frac{1}{\lambda+3} \, 
\biggl( 
1 + \lambda\cos^2\theta 
+ \mu \sin 2\theta  \cos\phi 
+ \frac{\nu}{2} \sin^2\theta  \cos 2\phi \nonumber\\
&+& \tau \sin\theta  \cos\phi
+ \eta \cos\theta  
+ \xi \sin^2\theta \sin 2\phi 
+ \zeta \sin 2\theta \sin\phi 
+ \chi \sin\theta  \sin\phi  
\biggr) \,. 
\en 
The relations between the three sets of hadronic structure functions are
shown in Appendix~\ref{app:Str_Func}. 

One of the interesting relations between the angular coefficients 
is the so-called Lam-Tung (LT) relation~\cite{Lam:1978zr}, which
was originally discovered in the naive parton model~\cite{Lam:1978zr}
and then confirmed in the collinear factorization approach at
order ${\cal O}(\alpha_s)$ 
(see Refs.~\cite{Collins:1984kg,Boer:2006eq,Berger:2007jw,%
Peng:2015spa}). The essence of the LT relation is that
the difference of the $A_0$ and $A_2$  angular
coefficients is equal to zero, i.e. the LT combination $A_{\rm LT} = A_0 - A_2$
vanishes in the parton model. Note, the LT relation
is also not affected by leading-order (LO) QCD
corrections. On the other hand,
at this order the $A_0$ and $A_2$ coefficients on magnitude are
equal to $A_0 = A_2 = \rho^2/(1+\rho^2)$, where $\rho^2 = Q_T^2/Q^2$, 
and therefore they vanish at small $Q_T$ limit. 
A violation of the LT relation, i.e. $A_{\rm LT} \neq 0$ occurs starting
with the second order in the $\alpha_s$ expansion~\cite{Gauld:2017tww}.
The explanation of the violation of the LT relation has been done
in Ref.~\cite{Peng:2015spa}. This phenomena was related to
a presence of a nonzero component of the quark-antiquark axis in the
direction normal to the plane of colliding hadrons. Such a noncomplanarity
between the partonic and hadonic planes in the rest frame of the
gauge boson occurs starting the second order in the $\alpha_s$,
when two or more gluons are radiated.

The angular coefficient $A_4$ is related to another important
quantity, the forward-backward (FB) asymmetry $A_{\rm FB}$.
In particular, the FB asymmetry $A_{\rm FB}$ is the property of 
the DY angular $\cos{\theta}$ distribution in the Collins-Soper frame
\eq
\rho_N = 
\frac{dN}{d \cos{\theta}}
= \frac{1}{2 \pi}
\, \int\limits_0^{2 \pi} d\phi \, \frac{dN}{d\Omega} 
= \frac{3}{16\pi}
\Bigg[1 + \cos^2\theta+\frac{A_0}{2}(1-3 \cos^2\theta)
+ A_4 \cos\theta\Bigg] \,. 
\en 
The latter is integrated
in the forward $(+)$ and backward $(-)$ directions
\eq
N_\pm =  \pm \int\limits^{\pm 1}_0  d\cos\theta \ \rho_N 
\en 
leading to the quantity of the interest, $A_{\rm FB}$,
which is also expressed though $A_4$ as 
\eq
A_{\rm FB} = \dfrac{N_+ - N_-}{N_+ + N_-}
= \frac{3}{8} \, A_4 \, .
\label{A-fb}
\en
In Refs.~\cite{Gutsche:2015mxa,Ivanov:2015tru}, 
the quantity of {\it convexity} was proposed 
parametrizing the $\cos^2\theta$ term in the angular
distributions of the exclusive decays of heavy hadrons. 
In particular, it was suggested to isolate the $\cos^2\theta$ term
from the linear $\cos\theta$ term by taking the second derivative of
angular distribution with respect to $\cos\theta$. We propose to derive
the quantity convexity $A_{\rm conv}$ relevant for the DY process
following the idea of Refs.~\cite{Gutsche:2015mxa,Ivanov:2015tru}:
\eq
A_{\rm conv} = \dfrac{\rho_N^{(2)}}{N_+ + N_-}\,,
\en
where $\rho_N^{(2)} = d^2\rho_N/(d\cos\theta)^2$. 
One can see that the convexity $A_{\rm conv}$ is
related to the asymmetry parameter $A_0$ and
it parametrizes the $W_T - W_L$ asymmetry of the transverse and longitudinal
hadronic structure functions in terms of the parameter
$\lambda = (W_T - W_L)/(W_T + W_L)$ defined above in Eq.~(\ref{dNGreek})
and see also Appendix~\ref{app:Str_Func}: 
\eq
A_{\rm conv} = \frac{3}{8} \, (2 - 3 A_0) = \dfrac{3 \lambda}{3 + \lambda}
\,.
\en

\section{Perturbative results}
\label{pert_calc}

Hadronic structure functions $W(x_1,x_2)$ 
characterizing DY process with colliding
hadrons $H_1$ and $H_2$ 
are related to partonic-level structure functions
$w^{ab}(x_1,x_2)$ by the QCD collinear 
factorization formula~\cite{Lyubovitskij:2024civ} 
\eq\label{factorization}
W(x_1,x_2) =
\frac{1}{x_1 x_2} \, \sum\limits_{a,b}
\, \int\limits_{x_1}^1 \, dz_1
\, \int\limits_{x_2}^1 \, dz_2
\ w^{ab}(z_1,z_2) 
\, f_{a/H_1}\Big(\frac{x_1}{z_1}\Big) 
\, f_{b/H_2}\Big(\frac{x_2}{z_2}\Big) \,,
\en
where $f_{a/H}(\xi)$ with $\xi = x_1/z_1$
is the PDF describing the collinear $\xi$ distribution of
partons of type $a$ in a hadron $H$.

For our calculation of the $T$-even structure functions we use a convenient  
orthogonal basis of vectors $P,R,K$~\cite{Lyubovitskij:2021ges}, defined by
\eq\label{basis_PRK}
P^\mu &=& (p_1 + p_2)^\mu \,, 
\nonumber\\
R^\mu &=& (p_1 - p_2)^\mu \,,
\nonumber\\
K^\mu &=& k_1^\mu
      - P^\mu  \, \frac{P \cdot k_1}{P^2}
      - R^\mu  \, \frac{R \cdot k_1}{R^2}
      = - q^\mu
      + P^\mu  \, \frac{P \cdot q}{P^2}
      + R^\mu  \, \frac{R \cdot q}{R^2}\,,
\en
which obey the conditions 
\eq
P^2 = - R^2 = \hat{s}\,,
\quad
K^2 = - \frac{\hat{u} \hat{t}}{\hat{s}}\,,
\quad
P^2 R^2 K^2 =  \hat{s} \hat{t} \hat{u}
\,,
\quad 
P \cdot R = P \cdot K = R \cdot K = 0  \,. 
\en
Here $p_1$, $p_2$, and $k_1$ are the momenta of the two initial partons and
the final-state parton, respectively,
satisfying the momentum conservation relation
$p_1 + p_2 = k_1 + q$. Furthermore, 
$\hat{s} = (p_1+p_2)^2$,
$\hat{t} = (p_1-q)^2$,
$\hat{u} = (p_2-q)^2$, with $\hat{s} + \hat{t} + \hat{u} = Q^2$ the parton-level 
Mandelstam variables. 

The $(P,R,K)$ and $(T,X,Y,Z)$ bases are related by 
\eq 
X^\mu &=& \frac{T^\mu \, \sqrt{1+\rho^2}}{\rho}
- \frac{P^\mu z_{12}^{+}  + R^\mu z_{12}^{-}}
{2 Q \rho \sqrt{1+\rho^2}} 
\,=\, \frac{\rho \Big(P^\mu z_{12}^{+}  + R^\mu z_{12}^{-}\Big)}
    {2 Q \sqrt{1+\rho^2}} - \frac{K^\mu \, \sqrt{1+\rho^2}}{Q \rho}
\,, \nonumber\\[2mm]   
Z^\mu &=& \frac{P^\mu z_{12}^{-}  + R^\mu z_{12}^{+}}
{2 Q \sqrt{1+\rho^2}}
\,, \nonumber\\[2mm]      
Y^\mu &=& - \, \epsilon^{\mu PRK}  \frac{z_1 z_2}{Q^3 \rho (1+\rho^2)} \,,
\en
where $z_{12}^{\pm} = z_1 \pm z_2$, $Q = \sqrt{Q^2}$, and
$\epsilon^{\mu PRK} = \epsilon^{\mu\nu\alpha\beta} 
\, P_\nu \, R_\alpha \, K_\beta$.

Also we will use the perpendicular $D$-dimensional metric tensor 
$g^{\mu\nu}_\perp$ introduced in Ref.~\cite{Lyubovitskij:2021ges}
\eq\label{kinematics2}  
g^{\mu\nu}_\perp
= g^{\mu\nu} 
- \frac{P^\mu P^\nu}{P^2}
- \frac{R^\mu R^\nu}{R^2}
- \frac{K^\mu K^\nu}{K^2}  \,, 
\en
which obeys the conditions
$g^{\mu\nu}_\perp \, V_\mu = 0$ with $V = P, R, K$
and $g^{\mu\nu}_\perp \, g_{\mu\nu; \perp} = D - 3$. 

We may project onto the parton-level $T$-even structure functions 
using the following relations:
\eq
w_{T} &=& \frac{1}{2} (X^\mu X^\nu + Y^\mu Y^\nu) \, w_{\mu\nu}
\nonumber\\
&=& \frac{1}{2} \,
\biggl[  \frac{\Big(P^\mu z_{12}^{+} + R^\mu z_{12}^{-} \Big)
         \,    \Big(P^\nu z_{12}^{+} + R^\nu z_{12}^{-} \Big)}
         {4 Q^2 \, \rho^2 \, (1+\rho^2)}
       - g^{\mu\nu}_\perp \biggr]  \, w_{\mu\nu} \,, \nonumber\\[2mm]   
w_{L} &=& Z^\mu Z^\nu \, w_{\mu\nu}
\nonumber\\
&=& \frac{\Big(P^\mu z_{12}^{-} + R^\mu z_{12}^{+} \Big)
    \,    \Big(P^\nu z_{12}^{-} + R^\nu z_{12}^{+} \Big)} 
         {4 Q^2 \, (1+\rho^2)}  \, w_{\mu\nu} \,, \nn
\nonumber\\[2mm]   
w_{\Delta\Delta} &=& \frac{1}{2} (Y^\mu Y^\nu - X^\mu X^\nu) \, w_{\mu\nu} 
\nonumber\\
&=& - \frac{1}{2} \,
\biggl[\frac{\Big(P^\mu z_{12}^{+} + R^\mu z_{12}^{-} \Big)
         \,    \Big(P^\nu z_{12}^{+} + R^\nu z_{12}^{-} \Big)} 
         {4 Q^2 \, \rho^2 \, (1+\rho^2)}
       + g^{\mu\nu}_\perp \biggr]  \, w_{\mu\nu} \,,
\nonumber\\[2mm]   
w_{\Delta} &=& - \frac{1}{2} (X^\mu Z^\nu + Z^\mu X^\nu) \, w_{\mu\nu} 
\nonumber\\
&=& 
\frac{1}{4 Q^2 \, \rho \, (1+\rho^2)}  \,
\biggl[\Big(P^\mu P^\nu + R^\mu R^\nu\Big) \, (z_1^2 - z_2^2)
  + \Big(P^\mu R^\nu + P^\nu R^\mu\Big) \, (z_1^2 + z_2^2)
\biggr]  \, w_{\mu\nu} \,, 
\nonumber
\en
\eq
w_{T_P} &=& - \frac{i}{2} (X^\mu Y^\nu - Y^\mu X^\nu) \, w_{\mu\nu}
\nonumber\\
&=& \frac{i z_1 z_2}{4 Q^4 \, \rho^2 \, (1+\rho^2)^{3/2}}  \, 
\biggl[ \epsilon^{\mu PRK} \Big(P^\nu z_{12}^{+} + R^\nu z_{12}^{-} \Big)
      - \epsilon^{\nu PRK} \Big(P^\mu z_{12}^{+} + R^\mu z_{12}^{-} \Big)
  \biggr]  \, w_{\mu\nu} \,, \nn
\nonumber\\[2mm]   
w_{\nabla_P} &=& - \frac{i}{2} (Y^\mu Z^\nu - Z^\mu Y^\nu) \, w_{\mu\nu}\ \nn
\nonumber\\
&=& \frac{i z_1 z_2}{4 Q^4 \, \rho \, (1+\rho^2)^{3/2}}  \, 
\biggl[ \epsilon^{\mu PRK} \Big(P^\nu z_{12}^{-} + R^\nu z_{12}^{+} \Big)
      - \epsilon^{\nu PRK} \Big(P^\mu z_{12}^{-} + R^\mu z_{12}^{+} \Big)
  \biggr]  \, w_{\mu\nu} \, 
\,. 
\label{W_projects}
\en

\begin{figure}[b]
  
  \includegraphics[height=3cm,trim={2cm 23.5cm 12cm 2cm},clip]
                  {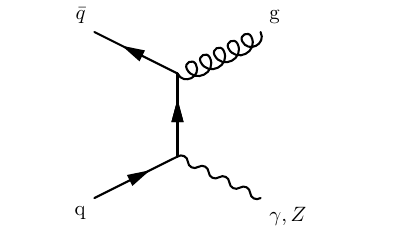}
	          \includegraphics[height=3cm,trim={2cm 23.5cm 12cm 2cm},clip]
                  {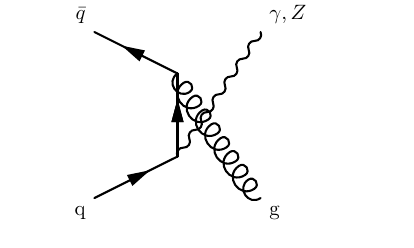}  
	\caption{Partonic-level quark-antiquark annihilation diagrams
         contributing to the DY cross section at order $\alpha_s$.}
	\label{Diag_qq}

\vspace*{.4cm}	        

\includegraphics[height=3cm,trim={2cm 23.5cm 12cm 2cm},clip]
                {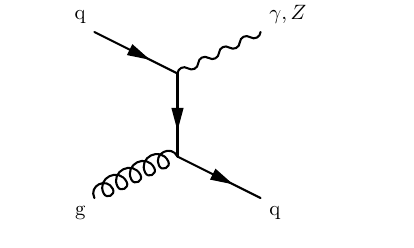}
	        \includegraphics[height=3.2cm,trim={2cm 23.5cm 12cm 2cm},clip]
                {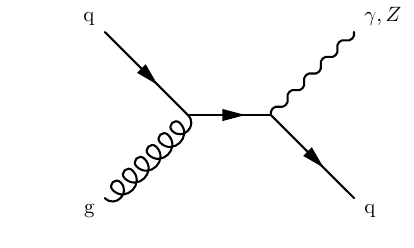}  
        \caption{Partonic-level quark-gluon Compton scattering diagrams
        contributing to the DY cross section at order $\alpha_s$.}
	\label{Diag_qg}	
\end{figure}

It is known that in case of the DY processes the $Q_T$ dependence of
the partonic structure functions starts at order ${\cal O}(\alpha_s)$
in the $\alpha_s$ expansion of the angular distributions.
At this order there are two types 
of subprocesses at the partonic level, which contribute: 
(a) quark-antiquark annihilation (Fig.~\ref{Diag_qq}) and 
(b) Compton quark-gluon scattering (Fig.~\ref{Diag_qg}).
Also we should take into account the subprocesses, where quarks are replaced 
by antiquarks.
We will comment on their contribution to the structure functions later.  

\vspace*{.5cm}

Before we display the results for the partonic structure functions, we
should specify the electroweak couplings, which occur in this quantities. 
First, we define the QCD color factors
\eq
C_F = \frac{N_c^2-1}{2 N_c} = \frac{4}{3}\,, \qquad
C_A = N_c = 3  \,, \qquad 
T_F = \frac{1}{2}\,, 
\en
which at large $N_c$ scale 
as ${\cal  O}(N_c)$, ${\cal  O}(N_c)$, ${\cal  O}(1)$, respectively.

The color factors that contribute to the partonic
subprocesses of quark-antiquark annihilation ($C_{q\bar q}$)  
and quark-gluon scattering $(C_{qg})$ are as follows: 
\eq
C_{q\bar q} = \frac{C_F}{N_c} = \frac{N_c^2-1}{2 N_c^2} = \frac{4}{9}
\,, \quad 
C_{qg} &=& \frac{T_F}{N_c} = \frac{1}{2 N_c} = \frac{1}{6} \,. 
\en 
The specific couplings, which occur in the
partonic structure functions are 
\eq
g_{q\bar q; i} = (8 \, \pi^2 \e_q^2 \, \alpha_s) \, 
C_{q\bar q} \, g_{{\rm EW}; i}^{Z\gamma/W} = G_i C_{q\bar q}
\en
and
\eq
g_{qg; i} =  (8 \, \pi^2 \e_q^2 \, \alpha_s) \,
C_{qg} \,  g_{{\rm EW}; i}^{Z\gamma/W}= G_i C_{q g}
\,, 
\en
where the index $i=1,2$ corresponds to two different cases.
In particular, the couplings with index $i=1$ are relevant
for the calculation of the $P$-even $W_T$, $W_L$, $W_{\Delta\Delta}$,
and $W_\Delta$ structure functions, while the couplings with index $i=2$
are used in the calculation of the $P$-odd structure functions
$W_{T_P}$ and $W_{\nabla_P}$. 
In the above equations the subscripts ${q\bar q}$ and ${q g}$ indicate
the specific partonic subprocesses, $e_q$ is the quark electric charge
of flavor $q$, $g_{{\rm EW}; 1}$ and $g_{{\rm EW}; 2}$ are the specific 
electroweak couplings including the products of couplings of the gauge 
bosons ($W^\pm$, $Z^0$, $\gamma$) with quarks and leptons. 
In the case of neutral gauge bosons  $Z^0$ and $\gamma$
we take into account their interference. In particular,
for the calculation of the $T$-even structure functions
we need the following couplings
\eq
g_{\rm EW; 1}^{Z\gamma} &=& 1 + 2 \, g^V_{Zq} \, g^V_{Zl} \, \mathrm{Re}[D_Z(Q^2)]
+ \Big((g_{Zq}^{V})^2 + (g_{Zq}^{A})^2\Big)  \Big((g_{Z\ell}^{V})^2
+ (g_{Z\ell}^{A})^2\Big) \,|D_Z(Q^2)|^2\,,
\nonumber\\[2mm] 
g^{Z\gamma}_{\rm EW; 2}&=&2 \, g_{Zq}^{A} \, \Big[
2 \, g_{Zq}^{V}  \, \Big(g_{Z\ell}^{A} \, g_{Z\ell}^{V}\Big) \, |D_Z(Q^2)|^2
+ g_{Z\ell}^{A} \, \mathrm{Re}[D_Z(Q^2)]\Big]  
\en
in case of the $(Z^0$, $\gamma$) bosons
and
\eq
g^{W}_{\rm EW; 1} &=& \Big((g_{Wqq'}^{V})^2  + (g_{Wqq'}^{A})^2\Big)
                 \, \Big((g_{W\ell}^{V})^2 + (g_{W\ell}^{A})^2\Big)
\, |V_{qq'}|^2 \, |D_W(Q^2)|^2\,,
\nonumber\\[2mm] 
g^{W}_{\rm EW; 2} &=& 4 \, \Big(g_{Wqq'}^{A} \,
g_{Wqq'}^{V}\Big) \, \Big(g_{W\ell}^{A}  \, g_{W\ell}^{V}\Big)  
\, |V_{qq'}|^2  \, |D_W(Q^2)|^2 
\en
in the case of the $W^\pm$ gauge bosons, 
where 
\eq
g_{W\ell}^{V} &=& g_{W\ell}^{A}
\, = \, g_{Wqq'}^{V} \, = \, g_{Wqq'}^{A}
\, = \,  \frac{1}{2 \sin\theta_W \sqrt{2}}
\,,
\nonumber\\[2mm] 
g_{Z\ell}^{V} &=& - \frac{1 - 4 \sin^2\theta_W}{2 \sin 2\theta_W}
\,,
\quad \hspace*{.1cm} 
g_{Z\ell}^{A} \, = \, - \frac{1}{2 \sin 2\theta_W}
\,,
\nonumber\\[2mm] 
g_{Zu}^{V} &=& \frac{1 - 8/3 \sin^2\theta_W}{2 e_q \sin 2\theta_W}
\,,
\quad 
g_{Zd}^{V} \, = \, - \frac{1 - 4/3 \sin^2\theta_W}{2 e_q \sin 2\theta_W}
\,,
\nonumber\\[2mm] 
g_{Zu}^{A} &=&   \frac{1}{2 e_q \sin 2\theta_W}
\,,
\quad \hspace*{.6cm} 
g_{Zd}^{A} \, = \, - \frac{1}{2 e_q \sin 2\theta_W}    
\en
are the couplings of the weak gauge bosons with leptons,
up ($u$), and down ($d$) quarks normalized by electric charge
of lepton $e$ and quark $e_q$, respectively. The gauge couplings
$g$ and $g'$ of the electroweak theory are related with electric
charge $e$ accordingly: $e = g \sin\theta_W = g' \cos\theta_W$,
where $\theta_W$ is the Weinberg angle. One should stress
that $g_{\rm EW; 1}^W \equiv g_{\rm EW; 2}^W =
|V_{qq'}|^2 \, |D_W(Q^2)|^2/(16 \, \sin^4\theta_W)$.  
Here,
\eq
\mathrm{Re}[D_G(Q^2)] &=& 
\frac{(M_G^2-Q^2) Q^2}{(M_G^2-Q^2)^2 + M_G^2 \Gamma_G^2} \,, 
\nonumber\\[2mm] 
\mathrm{Im}[D_G(Q^2)] &=& 
\frac{M_G \Gamma_G Q^2}{(M_G^2-Q^2)^2 + M_G^2 \Gamma_G^2} 
\en
are the real and imaginary parts of the Breit-Wigner propagator
of the weak gauge boson $G=W^\pm, Z^0$, and $V_{qq'}$ is the element
of the Cabibbo-Kabayashi-Maskawa (CKM) matrix. 
Masses $M_{G}$ and total widths $\Gamma_{G}$ of weak gauge bosons 
are taken from Particle Data Group~\cite{Workman:2022ynf}; 
$M_{W^\pm} = 80.377 \pm 0.012$ GeV,
$M_{Z^0} = 91.1876 \pm 0.0021$ GeV,
$\Gamma_{W^\pm} = 2.085 \pm 0.042$ GeV,
$\Gamma_{Z^0} = 2.4955 \pm 0.0023$ GeV.
For $\sin^2\theta_W$ we use that $\sin^2\theta_W =1-M_W^2/M_Z^2$.

It was stressed in Ref.~\cite{Lyubovitskij:2024civ}
the partonic structure functions with a single massless parton
in the final state contain the delta function 
$\delta\big((\hat{s}+\hat{t}+\hat{u}-Q^2)/\hat{s}\big)$.
Therefore, for convenience it was proposed to rewrite the partonic 
tensor $w^{ab}(z_1,z_2,\rho^2)$ as~\cite{Lyubovitskij:2024civ} 
\eq
\label{phase_space} 
w^{ab}(z_1,z_2,\rho^2)\,=\,\tilde{w}^{ab}(z_1,z_2,\rho^2)
\,\delta\big((\hat{s}+\hat{t}+\hat{u}-Q^2)/\hat{s}\big)\,.
\en
As we mentioned before, detailed analysis of the partonic and hadronic
structure functions including small $Q_T$ expansion at order $O(\alpha_s)$
and focusing to the electromagnetic DY process 
has been performed before in Refs.~\cite{Boer:2006eq,Berger:2007jw}.
Here we extend it to the case of the electroweak DY.  
In particular, for the $q \bar q$ annihilation subprocess
the expressions for the partonic $T$-even structure functions read 
\eq
\tilde{w}_{T}^{q \bar q}
&=& g_{q\bar q; 1} 
\ \biggl(\frac{1}{2} + \frac{Q^2 \hat{s}}{\hat{u} \hat{t}}\biggr)
\ \frac{(Q^2-\hat{u})^2 + (Q^2-\hat{t})^2}{(Q^2-\hat{u}) \, (Q^2-\hat{t})} 
\,,
\nonumber\\[2mm] 
\tilde{w}_{L}^{q \bar q}
&=& 2 \tilde{w}_{\Delta\Delta}^{q \bar q} \, = \, 
g_{q\bar q; 1}
\ \frac{(Q^2-\hat{u})^2 + (Q^2-\hat{t})^2}{(Q^2-\hat{u})  \, (Q^2-\hat{t})}
\,,
\nonumber\\[2mm] 
\tilde{w}_{\Delta}^{q \bar q}
&=& g_{q\bar q; 1} 
\ \sqrt{\frac{Q^2 \hat{s}}{\hat{u} \hat{t}}}
\ \frac{(Q^2-\hat{u})^2 - (Q^2-\hat{t})^2}{(Q^2-\hat{u})  \, (Q^2-\hat{t})} 
\,,
\nonumber\\[2mm] 
\tilde{w}_{T_P}^{q \bar q}
&=& g_{q\bar q; 2}
\ \frac{Q^2 \hat{s}}{\hat{u} \hat{t}} 
\ \sqrt{1+\frac{\hat{u} \hat{t}}{Q^2 \hat{s}}}
\ \frac{(Q^2-\hat{u})^2 + (Q^2-\hat{t})^2}{(Q^2-\hat{u})  \, (Q^2-\hat{t})} 
\,,
\nonumber\\[2mm] 
\tilde{w}_{\nabla_P}^{q \bar q}
&=& g_{q\bar q; 2} 
\ \sqrt{\frac{Q^2 \hat{s}}{\hat{u} \hat{t}}}
\ \sqrt{1+\frac{\hat{u} \hat{t}}{Q^2 \hat{s}}} 
\ \frac{(Q^2-\hat{u})^2 - (Q^2-\hat{t})^2}{(Q^2-\hat{u})  \, (Q^2-\hat{t})}
\,.
\en
For the $qg$ Compton scattering subprocess one finds
\eq
\tilde{w}_{T}^{q g}
&=& g_{q g; 1} 
\ \biggl[- \frac{(Q^2-\hat{s})^2 + (Q^2-\hat{t})^2}{\hat{s} \hat{t}} \,
+ \frac{\hat{u}}{2 \hat{s}}    
\ \frac{(Q^2+\hat{s})^2 + (Q^2-\hat{t})^2}{(Q^2-\hat{u}) \, (Q^2-\hat{t})} 
\, 
\biggr]
\,,
\nonumber\\[2mm] 
\tilde{w}_{L}^{q g}
&=& 2 \tilde{w}_{\Delta\Delta}^{q g} \, = \, 
- g_{q g; 1}
\ \frac{\hat{u}}{2 \hat{s}}    
    \, \frac{(Q^2+\hat{s})^2 + (Q^2-\hat{t})^2}{(Q^2-\hat{u}) \, (Q^2-\hat{t})} 
\,,
\nonumber\\[2mm] 
\tilde{w}_{\Delta}^{q g}
&=& g_{q g; 1} 
\ \sqrt{\frac{Q^2 \hat{u}}{\hat{s} \hat{t}}}  
\ \frac{2 (Q^2-\hat{t})^2 - (Q^2-\hat{u})^2}
   {(Q^2-\hat{u})  \, (Q^2-\hat{t})} \,
\,,
\nonumber\\[2mm] 
\tilde{w}_{T_P}^{q g}
&=& g_{q g; 2}
\ \frac{Q^2}{\hat{t}}
\ \sqrt{1+\frac{\hat{u} \hat{t}}{Q^2 \hat{s}}}
\ \frac{2 Q^2 (\hat{t}-\hat{u}) - (Q^2-\hat{u})^2}
{(Q^2-\hat{u})  \, (Q^2-\hat{t})}  
\,,
\nonumber\\[2mm] 
\tilde{w}_{\nabla_P}^{q \bar q}
&=& g_{q g; 2}
\ \sqrt{\frac{Q^2 \hat{u}}{\hat{s} \hat{t}}}
\ \sqrt{1+\frac{\hat{u} \hat{t}}{Q^2 \hat{s}}}
\ \frac{2 \hat{u} (Q^2+\hat{s}) + (Q^2-\hat{u})^2}
{(Q^2-\hat{u})  \, (Q^2-\hat{t})}  \,.
\en
For study of small $Q_T$ behavior of the hadronic structure functions,
it is convenient to express them in terms of the
variables $z_1$, $z_2$, and $\rho^2 = Q_T^2/Q^2$.
Using the following relations
(see more details in Appendix~\ref{app:Kinematics}): 
\eq 
& &\hat{s} = \frac{Q^2 + Q_T^2}{z_1 z_2} = 
\frac{Q_T^2}{(1-z_1)(1-z_2)} \,,\nonumber\\[2mm] 
& &\hat{t} = Q^2 - \frac{Q^2 + Q_T^2}{z_1} 
= - \frac{Q_T^2}{1-z_2} \,,\nonumber\\[2mm] 
& &\hat{u} = Q^2 - \frac{Q^2 + Q_T^2}{z_2} 
= - \frac{Q_T^2}{1-z_1} \,,\nonumber\\[2mm] 
& &   \frac{\hat{t}}{\hat{s}} = z_1 - 1\,,
\quad \frac{\hat{u}}{\hat{s}} = z_2 - 1\,,
\quad \frac{Q^2}{\hat{s}} = z_1 + z_2 - 1
\,,\nonumber\\[2mm] 
& &   \frac{Q^2-\hat{s}}{\hat{s}} = z_1 + z_2 - 2\,,
\quad \frac{Q^2-\hat{t}}{\hat{s}} = z_2\,,
\quad \frac{Q^2-\hat{u}}{\hat{s}} = z_1 
\,,\nonumber\\[2mm] 
& &\frac{\hat{u} \hat{t}}{Q^2 \hat{s}} = \rho^2 \,,
\quad 
\frac{(Q^2-\hat{u}) (Q^2-\hat{t})}{Q^2 \hat{s}} = 1 + \rho^2 \,.
\en
we get for the $q \bar q$ annihilation
\eq
\label{qq_func1}
\tilde{w}_T^{q\bar q} &=& g_{q\bar q; 1} 
\, \frac{1}{\rho^2} \, 
\biggl(1 + \frac{\rho^2}{2} \biggr) 
\, \frac{z_1^2 + z_2^2}{z_1 z_2}  
\,,  \nn\\[1mm] 
\tilde{w}_L^{q\bar q} &=& 2 \,\tilde{w}_{\Delta\Delta}^{q\bar q}  
\ = \ g_{q\bar q; 1} 
\, \frac{z_1^2 + z_2^2}{z_1 z_2}  
\,, \nn\\[1mm] 
\tilde{w}_\Delta^{q\bar q} &=&
g_{q\bar q; 1} 
\,  \frac{1}{\rho} \,
\, \frac{z_1^2 - z_2^2}{z_1 z_2}  
\,,  \nn\\[1mm] 
\tilde{w}_{T_P}^{q\bar q} &=&
g_{q\bar q; 2} 
\,  \frac{\sqrt{1+\rho^2}}{\rho^2}  
\, \frac{z_1^2 + z_2^2}{z_1 z_2}
\,,  \nn\\[1mm] 
\tilde{w}_{\nabla_P}^{q\bar q}                                                 
&=& g_{q\bar q; 2}
\,  \frac{\sqrt{1+\rho^2}}{\rho}  
\, \frac{z_1^2 - z_2^2}{z_1 z_2}  \,. 
\en
One can see that the $\omega^{q\bar q}$
partonic structure functions obey the conditions
\eq
\tilde{w}_L^{q\bar q} &=& 2 \ \tilde{w}_{\Delta\Delta}^{q\bar q}
\, = \, \frac{\rho^2}{1+\rho^2/2} \ \tilde{w}_T^{q\bar q}
\, = \,
\biggl(\frac{g_{q\bar q; 1}}{g_{q\bar q; 2}}\biggr)
\ \frac{\rho^2}{\sqrt{1+\rho^2}} \ \tilde{w}_{T_P}^{q\bar q}
\,,  \nn\\[1mm]
\tilde{w}_\Delta^{q\bar q} &=&
\biggl(\frac{g_{q\bar q; 1}}{g_{q\bar q; 2}}\biggr)
\ \sqrt{1+\rho^2} \ \tilde{w}_{\nabla_P}^{q\bar q}  
\en
For the $q g$ subprocess we get,
\eq
\label{qq_func1z1z2}
\tilde{w}_T^{q g} &=& g_{q g; 1} \,  \frac{1}{\rho^2} \,
\frac{1-z_2}{z_1 z_2} \,
\biggl(z_2^2 + (1 - z_1 z_2)^2  + \rho^2
\Big(1 - \frac{z_1^2}{2} - z_1 z_2 (z_1 + z_2)\Big)\biggr)  
\,,  \nn\\[1mm] 
\tilde{w}_L^{q g} &=& 2 \, \tilde{w}_{\Delta\Delta}^{q g}
\ = \ g_{q g; 1} \, 
\frac{1-z_2}{z_1 z_2} \ \Big(z_2^2 + (z_1 + z_2)^2\Big) 
\,, \nn\\[2mm] 
\tilde{w}_\Delta^{q g} &=&
g_{q g; 1} \, \frac{1}{\rho} \,
\frac{1-z_2}{z_1 z_2} \ \Big(z_1^2 - 2 z_2^2\Big) 
\,,  \nn\\[1mm] 
\tilde{w}_{T_P}^{q g} &=&
g_{q g; 2} \,  \frac{\sqrt{1+\rho^2}}{\rho^2}  
\, \frac{1-z_2}{z_1 z_2}
\ \Big(z_2^2 + (1 - z_2)^2 - (1 - z_1)^2\Big)
\,,  \nn\\[1mm] 
\tilde{w}_{\nabla_P}^{q g}                                                 
&=& g_{q g; 2} \,  \frac{\sqrt{1+\rho^2}}{\rho}  
\, \frac{1-z_2}{z_1 z_2} \ 
\Big(1- 2 z_2^2 - (1-z_1)^2 + 2 z_2 (1 - z_1) \Big) \,.
\en

Next, following formalism of Ref.~\cite{Lyubovitskij:2024civ},
we substitute the phase space formula~(\ref{phase_space})
to the factorization formula~(\ref{factorization})
\eq\label{factorizationdelta}
W(x_1,x_2,\rho^2) =
\frac{1}{x_1 x_2} \, \sum\limits_{a,b}
\, \int\limits_{x_1}^1 \, dz_1
\, \int\limits_{x_2}^1 \, dz_2
\ \tilde{w}^{ab}(z_1,z_2,\rho^2)\,\delta\left((1-z_1)(1-z_2)
-\frac{\rho^2 \,  z_1 z_2}{1+\rho^2}\right) 
\, f_{a/H_1}\Big(\frac{x_1}{z_1}\Big) 
\, f_{b/H_2}\Big(\frac{x_2}{z_2}\Big) \,.\quad
\en
In Ref.~\cite{Lyubovitskij:2024civ} we extrapolated
hadronic structure functions to small values
of $Q_T$ by performing expansion in powers of $\rho^2 = Q_T^2/Q^2$
about $\rho=0$ and up to order $\rho^4$. Here we present exact analytical
result without restricting to specific order in $\rho^2$. More detailed
discussion of the small $Q_T$ expansion in the pQCD
can be found in Refs.~\cite{Lyubovitskij:2024civ,LVWZ:2023}. 

Finally, we discuss behavior of the hadronic structure functions 
under interchange of the partons in the colliding hadrons.
It leads to the interchange of the partonic momenta,
structure and distribution functions,
$z_1$ and $z_2$ variables as  
\eq\label{z1z2Interchange}
& &p_1 \leftrightarrow p_2\,,
\quad f_{a/H_1}\Big(\frac{x_1}{z_1}\Big) \leftrightarrow
      f_{b/H_2}\Big(\frac{x_2}{z_2}\Big) \,,
\quad z_1 \leftrightarrow z_2 \,,
\nonumber\\
& &
w_i(z_1,z_2) \leftrightarrow w_i(z_2,z_1)
\,, \quad i = T, L, \Delta\Delta, \nabla_P
\,,\nonumber\\
& &
w_i(z_1,z_2) \leftrightarrow - w_i(z_2,z_1) 
\,, \quad i = \Delta, T_P \,.
\en
Note, the total contributions to the hadronic structure functions
for each partonic subprocess (quark-antiquark or quark-gluon scattering)
include the sum with taking into account of interchange of partons in
two colliding hadrons. In particular, the corresponding sums
for the $q\bar{q}$ and $qg$ subprocesses read as 
$W^{q\bar{q}} + W^{\bar{q}q}$ and $W^{qg} + W^{gq}$.
Following interchange transformation rules~(\ref{z1z2Interchange})
we find that the total contributions to the $T$, $L$, $\Delta\Delta$,
and $\nabla_P$ hadronic structure functions are symmetric under
interchange of partons for both quark-antiquark and quark gluon subprocesses,
while the $\Delta$ and $T_P$ hadronic structure functions are antisymmetric. 
Symmetric and antisymmetric properties of structure function at interchange 
of partons are simple to see from Eq.~(\ref{W_projects}). 
One should mention that such a property of the $\Delta$ hadronic structure
function was discussed before in Ref.~\cite{Berger:2007jw}.

For the $i = T, L, \Delta\Delta$ hadronic structure functions
the total contributions are given by 
\eq
W^{q \bar q}_{i} + W^{\bar q q}_{i} &\propto&
\left[
  f_{q/H_1}\Big(\frac{x_1}{z_1}\Big) 
  f_{\bar{q}/H_2}\Big(\frac{x_2}{z_2}\Big)
+ f_{\bar{q}/H_1}\Big(\frac{x_1}{z_1}\Big)
  f_{q/H_2}\Big(\frac{x_2}{z_2}\Big)
  \right]  \ (z_1^2+z_2^2)\,,
\nonumber\\
W^{q g}_{i} + W^{g q}_{i} &\propto&
  f_{q/H_1}\Big(\frac{x_1}{z_1}\Big) \, 
  f_{g/H_2}\Big(\frac{x_2}{z_2}\Big) \ 
  v_i(z_1,z_2) 
+ f_{g/H_1}\Big(\frac{x_1}{z_1}\Big) \, 
  f_{q/H_2}\Big(\frac{x_2}{z_2}\Big) \ 
  v_i(z_2,z_1)\,,
\en
where 
\eq
v_T(z_1,z_2) &=& (1-z_2)
\biggl(z_2^2 + (1 - z_1 z_2)^2  + \rho^2
\Big(1 - \frac{z_1^2}{2} - z_1 z_2 (z_1 + z_2)\Big)\biggr)  
\,,  \nonumber\\[1mm]
v_L(z_1,z_2) &=& 2 \, v_{\Delta\Delta}(z_1,z_2)
= (1-z_2) \ \Big(z_2^2 + (z_1 + z_2)^2\Big) 
\,.  
\en 
For the $\nabla_P$ structure functions 
the total contributions read as
\eq
W^{q \bar q}_{\nabla_P} + W^{\bar q q}_{\nabla_P} &\propto&
\biggl[
  f_{q/H_1}\Big(\frac{x_1}{z_1}\Big) \, 
  f_{\bar{q}/H_2}\Big(\frac{x_2}{z_2}\Big)
- f_{\bar{q}/H_1}\Big(\frac{x_1}{z_1}\Big) \,
  f_{q/H_2}\Big(\frac{x_2}{z_2}\Big)
  \biggr]  \ (z_1^2-z_2^2)\,,
\nonumber\\
W^{q g}_{\nabla_P} + W^{g q}_{\nabla_P} &\propto&
f_{g/H_1}\Big(\frac{x_1}{z_1}\Big) \,
f_{q/H_2}\Big(\frac{x_2}{z_2}\Big) 
\ (1-z_2) \
\Big(1 - 2 z_2^2 - (1 - z_1)^2 + 2 z_2 (1 - z_1)\Big)
\nonumber\\
&+&
f_{q/H_1}\Big(\frac{x_1}{z_1}\Big) \,
f_{g/H_2}\Big(\frac{x_2}{z_2}\Big) 
\ (1-z_1) \
\Big(1 - 2 z_1^2 - (1 - z_2)^2 + 2 z_1 (1-z_2)\Big)
\,.
\en
For the total contributions to the $\Delta$ and $T_P$
hadronic structure functions we have 
\eq
W_\Delta^{q\bar{q}} + W_\Delta^{\bar{q}q}  &\propto&
\biggl[f_{q/H_1}\Big(\frac{x_1}{z_1}\Big) \,
       f_{\bar{q}/H_2}\Big(\frac{x_2}{z_2}\Big)
    +  f_{\bar{q}/H_1}\Big(\frac{x_1}{z_1}\Big) \, 
       f_{q/H_2}\Big(\frac{x_2}{z_2}\Big) \biggr] \
       (z_1^2-z_2^2) \,,
\nonumber\\
W_\Delta^{qg} + W_\Delta^{gq} &\propto&
       f_{q/H_1}\Big(\frac{x_1}{z_1}\Big) \,
       f_{g/H_2}\Big(\frac{x_2}{z_2}\Big) \
       (1-z_2) \ (z_1^2-2z_2^2)
\nonumber\\
&+&   f_{g/H_1}\Big(\frac{x_1}{z_1}\Big) \,
       f_{q/H_2}\Big(\frac{x_2}{z_2}\Big) \
       (1-z_1) \ (2z_1^2-z_2^2)
\en
and 
\eq 
W^{q\bar{q}}_{T_P} + W^{\bar{q}q}_{T_P} &\propto&
\biggl[f_{q/H_1}\Big(\frac{x_1}{z_1}\Big) \,
       f_{\bar{q}/H_2}\Big(\frac{x_2}{z_2}\Big)
     - f_{\bar{q}/H_1}\Big(\frac{x_1}{z_1}\Big) \,
       f_{q/H_2}\Big(\frac{x_2}{z_2}\Big)\biggr] \
       (z_1^2 + z_2^2)
\,, 
\nonumber\\
W^{qg}_{T_P} + W^{gq}_{T_P} &\propto&
  f_{q/H_1}\Big(\frac{x_1}{z_1}\Big) \, 
  f_{g/H_2}\Big(\frac{x_2}{z_2}\Big)
  \ (1-z_2) \ \Big(z_2^2+(1-z_2)^2-(1-z_1)^2\Big)
\nonumber\\
&-& f_{g/H_1}\Big(\frac{x_1}{z_1}\Big) \,
    f_{q/H_2}\Big(\frac{x_2}{z_2}\Big)
    \ (1-z_1) \ \Big(z_1^2+(1-z_1)^2-(1-z_2)^2\Big)
\,.
\en

\section{Small-$Q_T$ expansion}	
\label{small_Qt}

As we pointed out in Refs.~\cite{Lyubovitskij:2024civ,LVWZ:2023} 
in the small $Q_T$ expansion of hadronic structure functions
presented in Eq.~\eqref{factorizationdelta} we have three contributions: 
(1) the direct dependence of the partonic structure function on $Q_T$; 
(2) the phase space delta function has nontrivial $Q_T$ dependence; 
(3) the fraction variables $x_1$, $x_2$ have implicit $Q_T$ dependence.
Obviously, the first type of the contributions can be straightforwardly
taken into account by simple Taylor expansion of the partonic structure functions:
\eq\label{wabsmallQT}
\tilde{w}^{ab}(z_1,z_2,\rho^2) = \sum\limits_{n=0}^{\infty}
(\rho^2)^n \, \tilde{w}^{ab; (n)}(z_1,z_2) \,, 
\en
where $\tilde{w}^{ab; (n)}(z_1,z_2)$ is the $n$th order term
in the $\rho^2$ expansion of the partonic structure function given by
\eq
\tilde{w}^{ab; (n)}(z_1,z_2) = \frac{1}{n!} \,
\partial_{\rho^2}^n\tilde{w}^{ab}(z_1,z_2,\rho^2)\Big|_{\rho^2=0} 
\,. 
\en
The expansion of the second and third contributions discussed in detail
in Refs.~\cite{Lyubovitskij:2024civ,LVWZ:2023}. 

The small $Q_T$ expansion of the phase space delta function was extensively
discussed in literature (see, e.g.,
Refs.~\cite{Meng:1995yn,Boer:2006eq,Berger:2007jw,Lyubovitskij:2024civ,LVWZ:2023}). 
In particular, its expansion to leading order
$\mathcal{O}(\rho^2)$ reads~\cite{Meng:1995yn,Boer:2006eq,Berger:2007jw,%
Lyubovitskij:2024civ,LVWZ:2023}, 
\eq
\label{delta_QT}
\delta\left((1-z_1)(1-z_2)-\frac{\rho^2}{1+\rho^2} z_1 z_2\right)
&=&\frac{\delta(1-z_1)}{(1-z_1)_+}+\frac{\delta(1-z_2)}{(1-z_2)_+}
\nonumber\\
&-&\delta(1-z_1)\delta(1-z_2)\log\rho^2+\mathcal{O}(\rho^2) \,,
\en
where the ``plus'' distribution $1/(1-z)_+$ is defined by 
\eq 
\int\limits_0^1 dz \,
\frac{f(z)}{(1-z)_+}\equiv\int_0^1 dz\,\frac{f(z)-f(1)}{1-z}\,,
\en
for a function $f(z)$ regular at $z=1$.
In Ref.~\cite{Gelfand} a general method for the expansion
of the integrals containing generalized functions (like delta-function)
was proposed and developed. It was based on the Mellin
integral techniques. Following these ideas,
in Ref.~\cite{LVWZ:2023} an algorithm for the small $Q_T$ expansion
of arbitrary singular function valid to arbitrary order of $\rho^2$
and arbitrary number of radiated partons have been formulated.

Here we present the exact formula for the small $Q_T$ expansion of the
delta function derived in two steps. First, we performed integration over
one of the variables $z_1$ or $z_2$ using delta function, e.g., over $z_2$ as:
\eq
I = \int\limits_{x_1}^1  dz_1 \int\limits_{x_2}^1 dz_2  \, 
\delta\left((1-z_1)(1-z_2)-\frac{\rho^2}{1+\rho^2} z_1 z_2 \right)
\,  \varphi(z_1,z_2) = (1 + \rho^2) 
\int\limits_{x_1}^{\sigma(x_2)} \, \frac{dz_1}{1 + \rho^2 - z_1} 
\,  \varphi(z_1,\sigma(z_1)) \,, 
\en
where $\sigma(x) = 1 - \frac{\rho^2 x}{1 + \rho^2 - x}$   
and $\varphi(z_1,\sigma(z_1))$ is a generic regular function. 
Second, in the remaining one-dimensional integral
we write the second argument in the function $\varphi(z_1,\sigma(z_1))$
as $z_2 = \sigma(z_1) = 1 + \rho^2
- \frac{\rho^2 (1 + \rho^2)}{1 + \rho^2 - z_1}$   
and make the Taylor-expansion of $\varphi(z_1,\sigma(z_1))$ around
$z_2 = 1 + \rho^2$: 
\eq
\int\limits_{x_1}^{\sigma(x_2)} \, dz_1
\,  \varphi(z_1,\sigma(z_1)) = 
(1 + \rho^2) \, \sum\limits_{N=0}^{\infty}
\frac{(- \rho^2 (1 + \rho^2))^N}{N!} 
\, \int\limits_{x_1}^{\sigma(x_2)} \, dz_1
\,  \frac{\varphi^{(N)}_{z_2}(z_1,1 + \rho^2)}{(1 + \rho^2 - z_1)^{N+1}} 
\,,
\en
where $\varphi^{(N)}_{z_2}(z_1,1 + \rho^2)
= \frac{\partial^{N}}{\partial z_2^{N}} \varphi(z_1,z_2)\bigg|_{z_2 = 1 + \rho^2}$.

After straightforward calculation we derive the desired
formula for the small $Q_T$ expansion of the delta function
up to arbitrary order in $\rho^2$,
\eq
I &=& \int\limits_{x_1}^1 dz_1 \int\limits_{x_2}^1 dz_2  \, 
\delta\left((1-z_1)(1-z_2)-\frac{\rho^2}{1+\rho^2} z_1 z_2 \right)
\,  \varphi(z_1,z_2)
\nonumber\\ 
&=& 
\int\limits_{x_1}^1 dz_1 \, \int\limits_{x_2}^1 dz_2 
\, \biggl(\delta(1-z_2)  \, G_1(z_1,z_2) + \delta(1-z_1) \, G_1(z_2,z_1)
\nonumber\\ 
&+& \delta(1-z_1) \, \delta(1-z_2) \, G_2(z_1,z_2)\biggr) \, \varphi(z_1,z_2) 
\,,
\en
where
\eq
G_1(z_1,z_2) &=&
\sum\limits_{N, M, K = 0}^{\infty} 
\, \frac{(\rho^2)^{N+M+K} \, (1 + \rho^2)^{N+1}}{(N!)^2  \, M! \, K!} \, 
\frac{(-1)^{N+K} \, (N+K)!}{(1-z_1)^{N+K+1}_{+,N+K}}  \, \partial_{z_2}^{N+M} \,,
\nonumber\\[2mm]
G_2(z_1,z_2) &=&
- \log\frac{\rho^2}{1 + \rho^2} \, \sum\limits_{N, M, K = 0}^{\infty} 
\, \frac{(\rho^2)^{N+M+K} \, (1 + \rho^2)^{N+1}}{(N!)^2  \, M! \, K!} \,
\partial_{z_1}^{N+K} \, \partial_{z_2}^{N+M}   \,.
\en 
In particular, if we restrict to the accuracy
${\cal O}(\rho^6,\rho^6\log\rho^2)$ as in Ref.~\cite{Lyubovitskij:2024civ},
then the expansion of the functions $G_1$ and $G_2$ reads	, 
\eq\label{small_QT_I}
G_1(z_1,z_2) &=& \frac{(1+\rho^2) (1+\rho^2\partial_{z_2}) +
  \rho^4\partial^2_{z_2}/2}{(1-z_1)_{+}}
- \frac{\rho^2 (1+\rho^2 + (1+3\rho^2) \partial_{z_2}
+ \rho^2 \partial^2_{z_2})}{(1-z_1)^2_{+,1}}
\nonumber\\
&+& \frac{\rho^4 (1+2 \partial_{z_2}+\partial^2_{z_2}/2)}{(1-z_1)^3_{+,2}}
+ {\cal O}(\rho^6,\rho^6\log\rho^2)
\,, \\
G_2(z_1,z_2) &=&
\rho^2 \biggl(1 + \rho^2 \Big(1/2+\partial_{z_1}+\partial_{z_2} 
+\partial_{z_1z_2}^2\Big)\biggr)
\nonumber\\
&-& \log\rho^2 \biggl(1 + \rho^2 (1-\rho^2)
   (1+\partial_{z_1}) (1+\partial_{z_2})
+ \rho^4  \Big(1 + 2 \partial_{z_1} + \partial^2_{z_1}/2\Big)
     \,  \Big(1 + 2 \partial_{z_2} + \partial^2_{z_2}/2\Big)
\biggr)
\nonumber\\
&+& {\cal O}(\rho^6,\rho^6\log\rho^2)
\,. 
\en
See details in Ref.~\cite{Lyubovitskij:2024civ}. 

Here  $1/(1-z)^m_{+,m-1}$ is a generalized plus distribution
of power $m$, defined by
\eq\label{PDF_d}
\int\limits_{x}^1 dz \, 
\frac{f(z)}{(1-z)^m_{+,m-1}}
&=& 
\int\limits_{x}^1 dz \, \biggl[\frac{1}{(1-z)^m_{+x,m-1}}
+ \delta(1-z) \, \log(1-x) \, \frac{(-1)^{m-1}}{(m-1)!} 
\, \partial_{z}^{m-1}
\nonumber\\
&-& \delta(1-z) \, 
\sum\limits_{j=2}^m
\, \frac{(-1)^{m-j}}{(j-1) \, (m-j)!} \, 
\biggl( \frac{1}{(1-x)^{j-1}} - 1 \biggr)
\, \partial_{z}^{m-j} 
\biggr] 
f(z) \,, 
\en
where $f(z)/(1-z)^m_{+x,m-1}$ is the $x$-plus distribution
\eq 
\int\limits_x^1 dz \frac{f(z)}{(1-z)^m_{+x,m-1}}\equiv
\int\limits_x^1 dz\frac{f(z)-\mathcal{T}^{m-1}_{z=1}f(z)}{(1-z)^m}\,,
\en  
derived by substraction from $f(z)$ its Taylor polynomial 
at $z=1$ to order $m-1$ 
\eq 
\mathcal{T}^{m-1}_{z=1}f(z) = 
\sum_{k=0}^{m-1}\frac{(-1)^k f^{(k)}(1)}{k!}\,(1-z)^k \,. 
\en
One should stress that our method is very simple and useful. 
In particular, it can be straightforwardly applied for 
the expansion of the phase space integrals: 
(1) for the small $Q_T$ expansion of 
delta functions occurring in other QCD processes,   
like semi-inclusive deep-inelastic scattering (SIDIS) and  
(2) for the small $Q_T$ expansion of  more complicated generalized functions,
like Heaviside $\theta$ function.

For example, the master integral for the SIDIS process involving delta
function is given by~\cite{Abele:2022spu}
\eq
I_{\rm SIDIS} &=& \int\limits_{x}^1 d\hat{x}
\int\limits_{z}^1 d\hat{z} \, 
\delta\biggl(R^2 \, \hat{x} \hat{z} - (1-\hat{x}) (1-\hat{z})\biggr)
\,  \varphi(\hat{x},\hat{z}) \,,
\en
where $x$ and $z$ are the Bjorken variables and the momentum fraction
variable that specifies the normalization of
outgoing hadron, respectively,
$\hat{x}$ and $\hat{z}$ are their partonic-level counterparts,
$R^2 = q_T^2/Q^2$ is the ratio of the square of the transverse gauge
boson momentum and Euclidean photon momentum squared.
We introduce a different notation for this ratio to
distinguish it from the DY ratio $\rho^2$. Comparing the delta
function occurring in the DY and SIDIS cases, we conclude
that the small $Q_T$ expansion in the SIDIS case can be
derived using the DY result upon substitution
$\rho^2 = R^2/(1-R^2)$. In the master integral $I_{\rm SIDIS}$
for simplicity we restrict to the regular
function $\varphi(\hat{x},\hat{z})$.

Taking into account above arguments the small $Q_T$ expansion is
given by
\eq
I_{\rm SIDIS} &=& \int\limits_{x}^1 d\hat{x} \int\limits_{z}^1 d\hat{z}  \, 
\, \biggl(\delta(1-\hat{z}) \, V_1(\hat{x},\hat{z})
        + \delta(1-\hat{x}) \, V_1(\hat{z},\hat{x}) 
\nonumber\\ 
&+& \delta(1-\hat{x}) \, \delta(1-\hat{z}) \, V_2(\hat{x},\hat{z})\biggr)
\, \varphi(\hat{x},\hat{z}) 
\,,
\en
where
\eq
V_1(\hat{x},\hat{z}) &=&
\sum\limits_{N, M, K = 0}^{\infty} 
\, \frac{(R^2)^{N+M+K}}{(1 - R^2)^{2N+M+K+1}}
\, \frac{1}{(N!)^2  \, M! \, K!} \, 
\frac{(-1)^{N+K} \, (N+K)!}{(1-\hat{x})^{N+K+1}_{+,N+K}}
\, \partial_{\hat{z}}^{N+M} \,,
\nonumber\\[2mm]
V_2(\hat{x},\hat{z}) &=& 
- \log R^2 \, \sum\limits_{N, M, K = 0}^{\infty}
\, \frac{(R^2)^{N+M+K}}{(1 - R^2)^{2N+M+K+1}}
\, \frac{1}{(N!)^2  \, M! \, K!} \,
\partial_{\hat{x}}^{N+K} \, \partial_{\hat{z}}^{N+M}   \,.
\en 
In particular, if we restrict to the accuracy
${\cal O}(R^6,R^6\log R^2)$, 
then the expansion of the functions $G_1$ and $G_2$ reads  
\eq\label{small_QT_SIDIS}
V_1(\hat{x},\hat{z}) &=& \frac{1}{(1-\hat{x})_{+}}
+ R^2 \, \hat{x} \, \frac{1 + \partial_{\hat{z}}}{(1-\hat{x})^2_{+,1}}
+ R^4  \, \hat{x}^2  \, \frac{
1 + 2 \partial_{\hat{z}}+\partial^2_{\hat{z}}/2}{(1-\hat{x})^3_{+,2}}
+ {\cal O}(R^6,R^6\log R^2)
\,, \\
V_2(\hat{x},\hat{z}) &=&
- \log R^2  \, \biggl(1 + R^2 (1+\partial_{\hat{x}}) (1+\partial_{\hat{z}})
+ R^4  \, 
(1+2 \partial_{\hat{x}}+\partial^2_{\hat{x}}/2)
(1+2 \partial_{\hat{z}}+\partial^2_{\hat{z}}/2)
\biggr)
\nonumber\\
&+& {\cal O}(R^6,R^6\log R^2)
\,. 
\en

As we stressed before, as an example of application
to other generalized functions 
we consider the small $Q_T$ expansion involving Heaviside $\theta$ function
in the DY process. The resulting formula reads, 
\eq
I_{\theta} &=& \int\limits_{x_1}^1  dz_1 \int\limits_{x_2}^1 dz_2  \, 
\theta\left((1-z_1)(1-z_2)-\frac{\rho^2}{1+\rho^2} z_1 z_2 \right)
\,  \varphi(z_1,z_2)
\nonumber\\ 
&=& 
\int\limits_{x_1}^1 dz_1 \, \int\limits_{x_2}^1 dz_2 
\, \biggl(1 + \delta(1-z_2)  \, F_1(z_1,z_2) + \delta(1-z_1) \, F_1(z_2,z_1)
\nonumber\\ 
&+& \delta(1-z_1) \, \delta(1-z_2) \, F_2(z_1,z_2)\biggr) \, \varphi(z_1,z_2) 
\,,
\en
where
\eq  
F_1(z_1,z_2) &=&
\sum\limits_{N, M = 0}^{\infty} 
\, (- 1)^N \,
\frac{(\rho^2)^{N+M+1}}{(N+1)!  \, M!} \,
\biggl(1 - \sum\limits_{K = 0}^{\infty} 
\, \frac{(-1)^{K} \, (N+K)!}{N! \, (K!)^2} \,
\frac{(\rho^2)^{K} \, (1 + \rho^2)^{N+1}}{(1-z_1)^{N+K+1}_{+,N+K}}
\biggr) \, \partial_{z_2}^{N+M}\,,
\nonumber\\[2mm]
F_2(z_1,z_2) &=&
\sum\limits_{N, M, K = 0}^{\infty} 
\, \frac{(\rho^2)^{N+M+K+1}}{(N+1)! \, N! \, M! \, K!} \,
\biggl(\log\frac{\rho^2}{1 + \rho^2} -
\frac{(1+\rho^2)^{N+1}-(\rho^2)^{N+1}}{N+1}\biggr) \,
\partial_{z_1}^{N+M} \, \partial_{z_2}^{N+K}
\nonumber\\[2mm]
&-&
\sum\limits_{N, M, K = 0}^{\infty}     \,
\sum\limits_{L = N + 1}^{\infty} (-)^L \, (\rho^2)^{M+K+L+1} \,
\frac{(1+\rho^2)^{N+1}-(\rho^2)^{N+1}}{(N+1)! \, (L+1)!  \, M! \, K!} \,
\, \Big(
    \partial_{z_1}^{N+K}   \, \partial_{z_2}^{L+M}
  + \partial_{z_1}^{L+K}   \, \partial_{z_2}^{N+M}
   \Big)
\,.
\en

Substituting the small-$Q_T$ expansion of the parton-level  
structure functions $w^{ab}(z_1,z_2,\rho^2)$ for the various 
partonic channels into Eq.~(\ref{factorization})  
we get the small-$Q_T$ expansion of  
the hadronic structure function $W(x_1,x_2,\rho^2)$ including  
two contributions discussed above (from the direct expansion of 
the partonic-level structure function $\tilde w^{ab}(z_1,z_2,\rho^2)$  
given by Eq.~(\ref{wabsmallQT}) and small $Q_T$ expansion of  
the phase space delta function) 
\eq\label{small_QT_II}
W_{\textrm{direct}+\delta}(x_1, x_2,\rho^2)
= \sum\limits_{i = 0}^{\infty} \, (\rho^2)^i \, 
W_i(x_1, x_2,L_\rho) 
\,,
\en
where following Ref.~\cite{Lyubovitskij:2024civ} we introduce the notation 
$L_\rho\equiv \log\rho^2$. It remains to take into account small $Q_T$
expansion due to implicit $Q_T$ dependence of the fraction variables
$x_1$ and $x_2$. 

The expansion coefficients $W_i(x_1, x_2,L_\rho)$ have the structure
\eq\label{Wi_general},
W_i(x_1, x_2,L_\rho) &=& \frac{1}{x_1x_2}\sum_{a,b}
\, \biggl[
   R_{ab,i}(x_1,x_2,L_\rho)
\, f_{a/H_1}(x_1)
\, f_{b/H_2}(x_2)
\nonumber\\
&+& \Big(P_{ba,i} \otimes f_{b/H_2}\Big)(x_2,x_1,L_\rho)  \ f_{a/H_1}(x_1)  
 +  \Big(P_{ab,i} \otimes f_{a/H_1}\Big)(x_1,x_2,L_\rho)  \ f_{b/H_2}(x_2) 
\biggr] 
\,,
\en 
where
\eq\label{convolution_general}
\big({\cal P} \otimes f\big)(x,y,L_\rho) = \int_x^1 
\, \frac{dz}{z} \, {\cal P}(z,y,L_\rho) \, f\Big(\frac{x}{z}\Big) 
\en  
denotes a generalized convolution, $R_i(x_1,x_2,L_\rho)$,
$P_{ba,i}(z_2,x_1,L_\rho)$, and $P_{ab,i}(z_1,x_2,L_\rho)$
are perturbative coefficient
functions containing differential operators acting on
the PDFs $f_{a/H_1}(x_1)$ and $f_{b/H_2}(x_2)$. 
We note that the generalized convolution~(\ref{convolution_general})
reverts to the ordinary one,
\eq\label{convolution_usual}
\big({\cal P} \otimes f\big)(x) = \int_x^1 
\, \frac{dz}{z} \, {\cal P}(z) \, f\Big(\frac{x}{z}\Big) \,,
\en 
when ${\cal P}(z,y,L_\rho)$ does not depend on $y$ and $L_\rho$.
Details are given in Appendix~\ref{app:Expansion}.
We stress that, as indicated in Eq.~(\ref{small_QT_II}),
the functions $W_i$ may carry dependence on $\log\rho^2$,
on top of the overall power of $\rho^2$ that they multiply.

However, Eq.~(\ref{small_QT_II}) is not yet the complete expansion.
As mentioned above, we need to take into account that $x_1$ and $x_2$ are
defined at finite $Q_T$ and hence must also be
expanded about their respective values at $Q_T = 0$, $x_1^0$ and
$x_2^0$.   
Therefore, we substitute $x_i = x_i^0 \sqrt{1+\rho^2}$ as arguments of
the structure functions $W_i$ and perform the $\rho^2$ expansions of
the latter.
We now present our final result for the full small-$Q_T$
expansion of the hadronic structure functions, including the
leading-power (LP) term $W^{\rm LP}(x_1^0, x_2^0,L_\rho)$, the
next-to-leading-power (NLP) term $W^{\rm NLP}(x_1^0, x_2^0,L_\rho)$, and the
next-next-to-leading-power (NNLP) term $W^{\rm NNLP}(x_1^0, x_2^0,L_\rho)$, etc.,
\eq\label{FinalFormulaQT} 
W(x_1, x_2,\rho^2) &=&
\sum\limits_{m = 0}^{\infty} \, (\rho^2)^m \, 
W^{{\rm N}^m{\rm LP}}(x_1^0, x_2^0,L_\rho) 
\nonumber\\[2mm]
&=&
W^{\rm LP}(x_1^0, x_2^0,L_\rho)
+ \rho^2 \, W^{\rm NLP}(x_1^0, x_2^0,L_\rho)
+ \rho^4 \, W^{\rm NNLP}(x_1^0, x_2^0,L_\rho) +  \ldots
\nonumber\\[2mm]
&=&
\sum\limits_{i = 0}^{\infty} \, 
\sum\limits_{s_1, s_2 = 0}^{\infty} \,
(\rho^2)^i \, (\sqrt{1+\rho^2}-1)^{s_1+s_2} 
\, \frac{(x_1^0)^{s_1} \, (x_2^0)^{s_2}}{s_1!  \, s_2!}
\, \partial_{x_1^0}^{s_1} \partial_{x_2^0}^{s_2} W_i(x_1^0,x_2^0,L_\rho)
\,,
\en
where $W^{{\rm N}^m{\rm LP}}(x_1^0, x_2^0,L_\rho)$ denotes 
the $i$th order term in the small $Q_T$ expansion of the structure
function including all three types of the $Q_T$ corrections
discussed in the beginning of this section. 
Here ${\rm N^0LP} =  {\rm LP}$, ${\rm N^1LP} =  {\rm NLP}$,
${\rm N^2LP} =  {\rm NNLP}$, etc. 
$\partial^{m}_{x_1} \partial^{n}_{x_2} W_i(x_1,x_2,L_\rho)$ denotes 
the $m$th partial derivative with respect to $x_1$ and the $n$th
partial derivative with respect to $x_2$.
The calculation techniques for taking these derivatives was discussed
in detail in Ref.~\cite{Lyubovitskij:2024civ}. 
In Appendix~\ref{app:Expansion} we present the complete formula for taking
this derivatives including all possible singularities
due to logarithms and $1/(1-z)$ poles. 

As we stressed above the expressions for the
$W^{{\rm N}^m{\rm LP}}(x_1^0, x_2^0,L_\rho)$ give the final and full results
(including all sources of the $Q_T$ corrections) for
the expansion of the hadronic structure functions
to desired order in the small $Q_T^2$ expansion. 
To get the analytic expression for any
$W^{{\rm N}^m{\rm LP}}(x_1^0, x_2^0,L_\rho)$ term one should make
the $i$th order partial derivative with respective $\rho^2$
without touching the nonanalytical logarithmic term $L_\rho$ 
using Eq.~(\ref{FinalFormulaQT}),
\eq
W^{{\rm N}^m{\rm LP}}(x_1^0, x_2^0,L_\rho) &=&
\frac{1}{m!} \, \frac{\partial^m}{\partial^m\rho^2}
W(x_1, x_2,\rho^2)\Big|_{\rho^2 = 0}
\nonumber\\
&=& \sum\limits_{i = 0}^{m}
\, \sum\limits_{s_1, s_2 = 0}^{s_1 + s_2 \le m - i}
\, \sum\limits_{k = 0}^{s_1 + s_2}
\, (-1)^{s_1 + s_2 - k}
\, C_{s_1 + s_2}^{k} \, C_{k/2}^{m-i}
\, \frac{(x_1^0)^{s_1} \, (x_2^0)^{s_2}}{s_1!  \, s_2!}
\, \partial_{x_1^0}^{s_1} \partial_{x_2^0}^{s_2} W_i(x_1^0,x_2^0,L_\rho)
\,. 
\en
In particular, the LP, NLP, and NNLP hadronic functions 
follow 
from the above expression and are given by~\cite{Lyubovitskij:2024civ} 
\eq
W^{\rm LP}(x_1^0, x_2^0,L_\rho) &=& W_0(x_1^0, x_2^0,L_\rho) \,,
\\
W^{\rm NLP}(x_1^0, x_2^0,L_\rho) &=& W_1(x_1^0, x_2^0,L_\rho)
+   \frac{1}{2} \biggl(
x_1^0 \ \partial_{x_1^0} W_0(x_1^0, x_2^0,L_\rho) 
    +   x_2^0 \ \partial_{x_2^0} W_{0}(x_1^0, x_2^0,L_\rho) 
    \biggr) \,, 
\\
W^{\rm NNLP}(x_1^0, x_2^0,L_\rho) &=& W_2(x_1^0, x_2^0,L_\rho)
+  \frac{1}{4} x_1^0 x_2^0
\ \partial_{x_1^0} \partial_{x_2^0} W_0(x_1^0, x_2^0,L_\rho) 
\nonumber\\
&-&   \frac{1}{8} \biggl(
    x_1^0    \ \partial_{x_1^0} W_0(x_1^0, x_2^0,L_\rho)
- 4 x_1^0    \ \partial_{x_1^0} W_1(x_1^0, x_2^0,L_\rho) 
-  (x_1^0)^2 \ \partial^2_{x_1^0} W_0(x_1^0, x_2^0,L_\rho) 
\biggr)
\nonumber\\  
&-&   \frac{1}{8} \biggl(
    x_2^0    \ \partial_{x_2^0}  W_0(x_1^0, x_2^0,L_\rho)
- 4 x_2^0    \ \partial_{x_2^0}  W_1(x_1^0, x_2^0,L_\rho) 
-  (x_2^0)^2 \ \partial^2_{x_2^0} W_0(x_1^0, x_2^0,L_\rho)
\biggr)  \,.
\en
The method for the calculation of the partial derivatives
$\partial_{x_1^0}^{s_1} \partial_{x_2^0}^{s_2} W_i(x_1^0,x_2^0,L_\rho)$
was proposed in Ref.~\cite{Lyubovitskij:2024civ}.
The main task here to calculate the terms containing
the convolution of the perturbative coefficient function and
the PDF. In Appendix~\ref{app:Expansion} we discuss a
generalization of the method proposed in Ref.~\cite{Lyubovitskij:2024civ}
to arbitrary perturbative coefficient function including
both possible logarithmic and pole endpoints $z \to 1$ singularities. 

Explicitly we obtain the following analytical results for
the LP contributions for the
$W_{J}^{{\rm LP}; a b}(x_1^0,x_2^0,L_\rho)$ to the $T$-even hadronic
structure functions (here $a b = q \bar q, q g$ and
$J = T, L, \Delta\Delta, \Delta, T_P, \nabla_P$),
\eq
\label{qq_limit}
W^{{\rm LP}; q \bar q}_{T}(x_1^0,x_2^0,L_\rho) &=& 
\frac{1}{\rho^2} \, 
W^{{\rm LP}; q \bar q}_{L}(x_1^0,x_2^0,L_\rho)  
\, = \, \frac{2}{\rho^2} \,
W^{{\rm LP}; q \bar q}_{\Delta\Delta}(x_1^0,x_2^0,L_\rho)
\nonumber\\
&=& \frac{g_{q\bar q; 1}}{g_{q\bar q; 2}} \, 
W^{{\rm LP}; q \bar q}_{T_P}(x_1^0,x_2^0,L_\rho)
\nonumber\\
&=& 
\frac{g_{q\bar q; 1}}{\rho^2 x_1^0 x_2^0} \, 
\frac{1}{C_F} \,
\biggl[- C_F (2 L_\rho + 3) \, q_1(x_1^0)  \, \bar q_2(x_2^0)
\nonumber\\
&+& q_1(x_1^0) \, \Big(P_{qq} \otimes \bar q_2\Big)(x_2^0) 
+ \Big(P_{qq} \otimes q_1\Big)(x_1^0) \, \bar q_2(x_2^0)  
\biggr]   \,, 
\\
W^{{\rm LP}; q \bar q}_{\Delta}(x_1^0,x_2^0,L_\rho)
&=& \frac{g_{q\bar q; 1}}{g_{q\bar q; 2}}  
\, 
W^{{\rm LP}; q \bar q}_{\nabla_P}(x_1^0,x_2^0,L_\rho)
\nonumber\\
&=& 
\frac{g_{q\bar q; 1}}{\rho x_1^0 x_2^0}  
\frac{1}{C_F} \,
\biggl[q_1(x_1^0) \, \Big(\tilde P_{qq} \otimes \bar q_2\Big)(x_2^0) 
- \Big(\tilde P_{qq} \otimes q_1\Big)(x_1^0) \, \bar q_2(x_2^0) 
\biggr]  \,, 
\en
for quark-antiquark annihilation process
and
\eq
\label{qg_limit}
W^{{\rm LP}; q g}_{T}(x_1^0,x_2^0,L_\rho)
&=& \frac{g_{q g; 1}}{g_{q g; 2}}  
\, 
W^{{\rm LP}; q g}_{T_P}(x_1^0,x_2^0,L_\rho)
\nonumber\\
&=& 
\frac{2 g_{q g; 1}}{\rho^2 x_1^0 x_2^0}  \, 
  q_1(x_1^0) \, \Big(P_{qg}^+ \otimes g_2\Big)(x_2^0) 
\,, 
\\
W^{{\rm LP}; q g}_{L}(x_1^0,x_2^0,L_\rho)  &=&
2 W^{{\rm LP}; q g}_{\Delta\Delta}(x_1^0,x_2^0,L_\rho)
\nonumber\\
&=& \frac{2 g_{q g; 1}}{x_1^0 x_2^0}  \, 
q_1(x_1^0) \, \Big(P_{qg}^- \otimes  g_2\Big)(x_2^0) 
\,, 
\\
W^{{\rm LP}; q g}_{\Delta}(x_1^0,x_2^0,L_\rho)
&=& \frac{g_{q g; 1}}{g_{q g; 2}}  
\, 
W^{{\rm LP}; q g}_{\nabla_P}(x_1^0,x_2^0,L_\rho)
\nonumber\\
&=&  \frac{2 g_{q g; 1}}{\rho x_1^0 x_2^0}  
\, q_1(x_1^0) \, \Big(\tilde P_{qg} \otimes  g_2\Big)(x_2^0) 
\en
for the quark-gluon Compton process, 
where we use the following notations for the
partonic splitting functions,
\eq
P_{qq}(z) &=& C_F \biggl[ \frac{1+z^2}{(1-z)_+} 
+ \frac{3}{2} \delta(1-z)\biggr] \,, \nonumber\\
P^{\pm}_{qg}(z)&=&T_F  \, [z^2 + (1 \mp z)^2] \,, \nonumber\\
\tilde P_{qq}(z)&=&C_F  \, [1 + z] \,, \nonumber\\
\tilde P_{qg}(z)&=&T_F  \, [1 - 2 z^2] \,. 
\en
We should stress that the LP hadronic structure functions
obey the following identities:
\eq
\label{qq_limit_ident}
W^{{\rm LP}; q \bar q}_{T}&=& 
\frac{1}{\rho^2} \, 
W^{{\rm LP}; q \bar q}_{L} 
\, = \, \frac{2}{\rho^2} \,
W^{{\rm LP}; q \bar q}_{\Delta\Delta} 
\, = \, \frac{g_{q\bar q; 1}}{g_{q\bar q; 2}} \, 
W^{{\rm LP}; q \bar q}_{T_P}
\,, \nonumber\\
W^{{\rm LP}; q \bar q}_{\Delta}
&=& \frac{g_{q\bar q; 1}}{g_{q\bar q; 2}}  
\, W^{{\rm LP}; q \bar q}_{\nabla_P}  
\en
and
\eq
\label{qg_limit2}
W^{{\rm LP}; q g}_{T}
&=& \frac{g_{q g; 1}}{g_{q g; 2}}  
\, W^{{\rm LP}; q g}_{T_P}
\,, 
\nonumber\\
W^{{\rm LP}; q g}_{L} &=&
2 W^{{\rm LP}; q g}_{\Delta\Delta}
\,,
\nonumber\\
W^{{\rm LP}; q g}_{\Delta}
&=& \frac{g_{q g; 1}}{g_{q g; 2}}  
\, 
W^{{\rm LP}; q g}_{\nabla_P}
\,.
\en
These identities are important and, in particular, to fix
the value of the angular coefficient $A_4$  at small $Q_T$. 
The coefficients $A_i$ (see definition in Appendix~\ref{app:Str_Func})
vanish in the limit $Q_T\to0$ except $A_4$ coefficient, 
because the LP $W_{T_P}^{\rm LP}$ structure function has the same
small $Q_T$ behavior as the transverse structure function $W_T^{\rm LP}$
The asymmetry coefficient $A_4$
is directly related to the FB asymmetry.
In Appendix~\ref{app:Expansion} we present analytical results for
the NLP structure functions. 

\section{Numerical analysis}
\label{Discussion}

In this section we discuss our results for the $T$-even angular coefficients
and compare our predictions with available data from the ATLAS and CMS
Collaborations at CERN. 

First, we illustrate the behavior of the hadronic structure functions
at different orders of the small $Q_T$ expansion.
In Ref.~\cite{Lyubovitskij:2024civ} we studied small $Q_T$ expansion
of the $T$-odd hadronic structure functions. As example, we 
considered the $q\bar{q}$ contribution to the hadronic double-flip
structure function, $W^{q \bar q}_{\Delta\Delta_P}(x_1,x_2)$. 
In particular, we compared the full expression without $Q_T$ expansion 
with the LP, NLP, and NNLP results. We used the CTEQ 6.1M PDFs of
Ref.~\cite{Stump:2003yu}, taken from LHAPDF~\cite{Buckley:2014ana},
along with their ManeParse~\cite{Clark:2016jgm}
Mathematica implementation. As representative of the kinematics
in the ATLAS measurements~\cite{ATLAS:2016rnf}
we chosen $\sqrt{s}=8$~TeV, $Q=100$~GeV, 
and the renormalization and factorization scales in the calculation 
are set to $\mu=\sqrt{Q^2+Q_T^2}$. We showed, that the LP piece 
describes the full result only at low $Q_T$ and rapidly  
departs from it for $Q_T>10$~GeV or $\rho^2 > 0.01$. 
Indeed, inclusion of the NLP term led to excellent
agreement with the full result out to $Q_T=40$~GeV ($\rho^2 = 0.16$),
only marginally further improved by the NNLP contribution. 
E.g., for $Q_T=20$~GeV, the LP result deviated from the full one by 
about 20\%, whereas at the NNLP the relative deviation is only $\sim 0.4\%$.

\begin{figure}[b]
        \includegraphics[height=5.25cm]{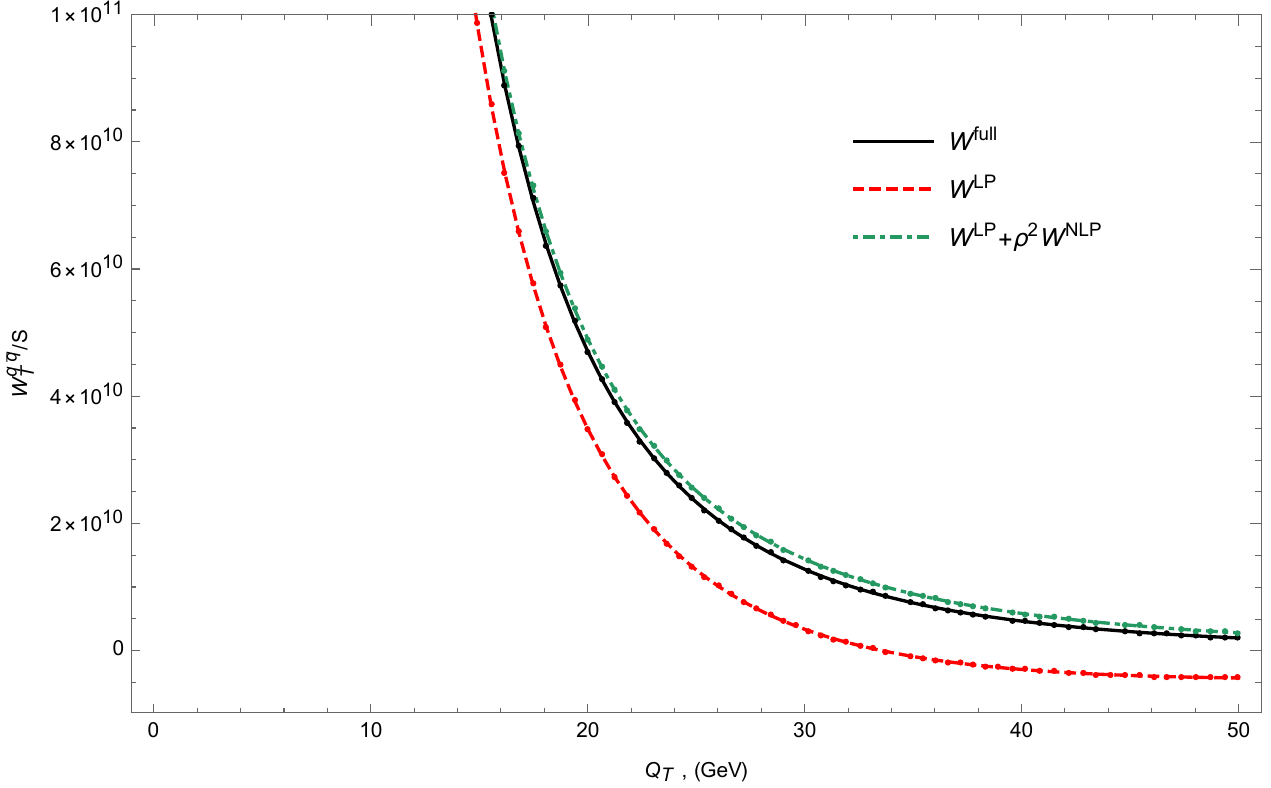}
        \qquad
        \includegraphics[height=5.25cm]{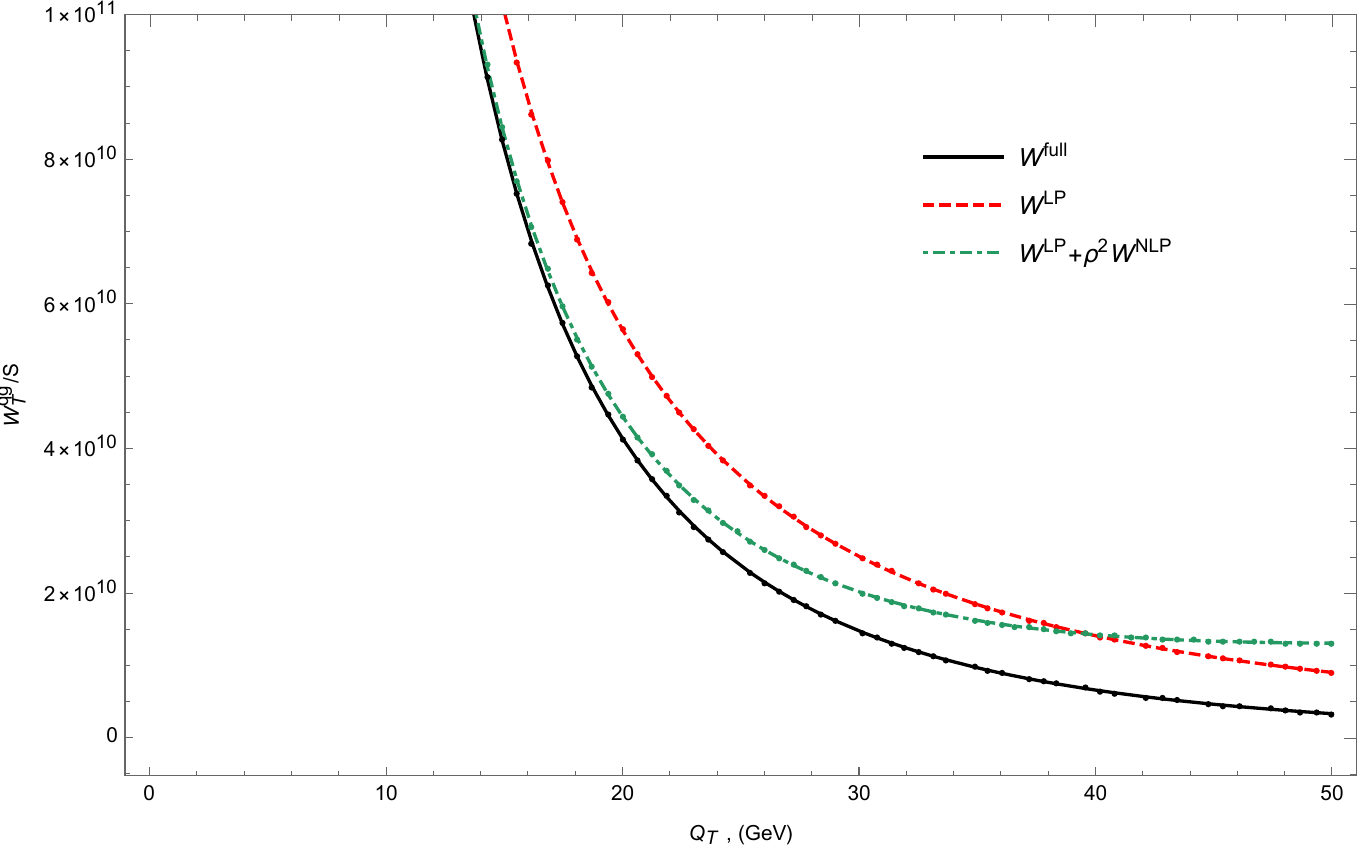}
        \caption{Comparison of the full analytical result for 
        the $W_T$ structure function (black solid line) 
        with expansions to LP (red dashed) and NLP (green dot-dashed)
        for two partonic subprocesses:
        (a) quark-antiquark scattering (left panel),
        (b) quark-gluon scattering (right panel). 
\label{WTsmallQT}}
\end{figure}

Here we present similar analysis restricting to the full result (without
small $Q_T$ expansion), LP and NLP contribution. As example, we consider
transverse structure function $W_T$. In Fig.~\ref{WTsmallQT} 
we present our results for the $Q_T$ dependence of the $W_T$ structure
function (total, LP, and NLP contributions) for the quark-antiquark
(left panel) and quark-gluon (right) panel subprocesses.
We consider the same kinematics ($\sqrt{s}=8$~TeV, $Q=100$~GeV) as in the
Ref.~\cite{Lyubovitskij:2024civ} and the ATLAS experiment~\cite{ATLAS:2016rnf}.
Also we use the CTEQ 6.1M PDFs of
Ref.~\cite{Stump:2003yu} from LHAPDF~\cite{Buckley:2014ana}
and ManeParse~\cite{Clark:2016jgm}. One can see that results for the $W_T$
hadronic structure function are similar to one obtained for the $T$-odd
hadronic structure functions. Again, the LP term is closed to the
full results only at low $Q_T$ and deviates from it at $Q_T>10$~GeV or
$\rho^2 > 0.01$. Inclusion of the NLP term gives good agreement with
the full result.

Second, we show a comparison of our predictions for the
$T$-even angular coefficients and data extracted by
the ATLAS Collaboration~\cite{ATLAS:2016rnf} for 
the eight angular coefficients $A_{i=0,..,7}$. 
The measurement was made in the $Z$-boson invariant mass window
$Q \in [80,100]$ GeV, as a function of $Q_T$.
ATLAS results for the angular coefficients were presented
for the case of integration over specific rapidity areas and
in two formats -- unregularized and regularized by bias analysis.
Besides, data presented for three areas of the rapidity $y$:
(a) $|y|<1$, (b) $1<|y|<2 $, and (c) $2<|y|<3.5$.

\begin{figure}[t]
	\includegraphics[height=6.5cm]{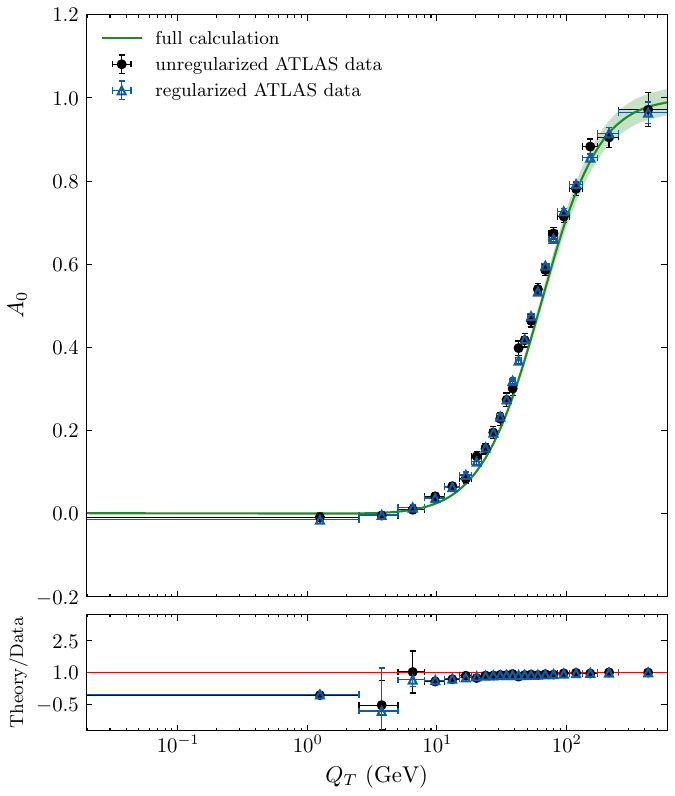}
	\includegraphics[height=6.5cm]{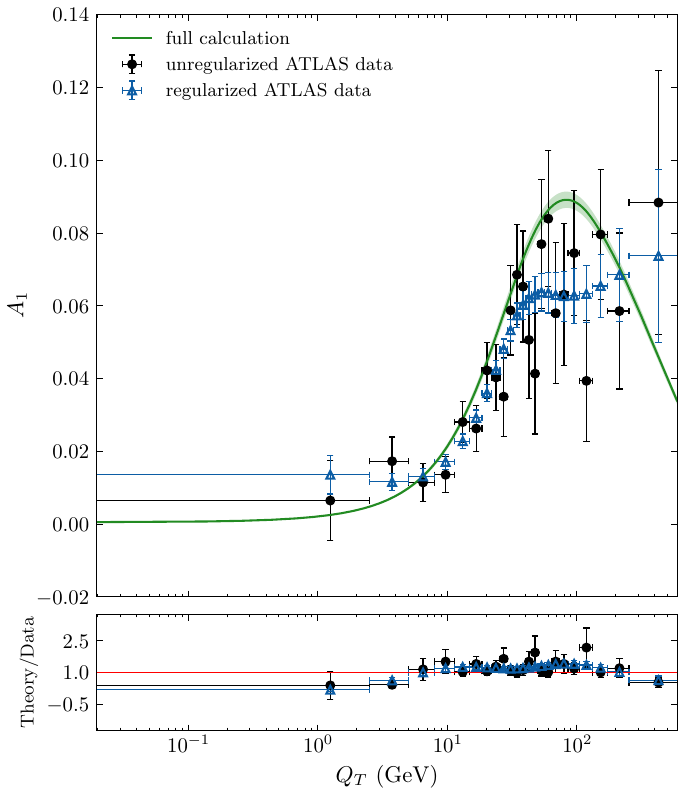}
	\includegraphics[height=6.5cm]{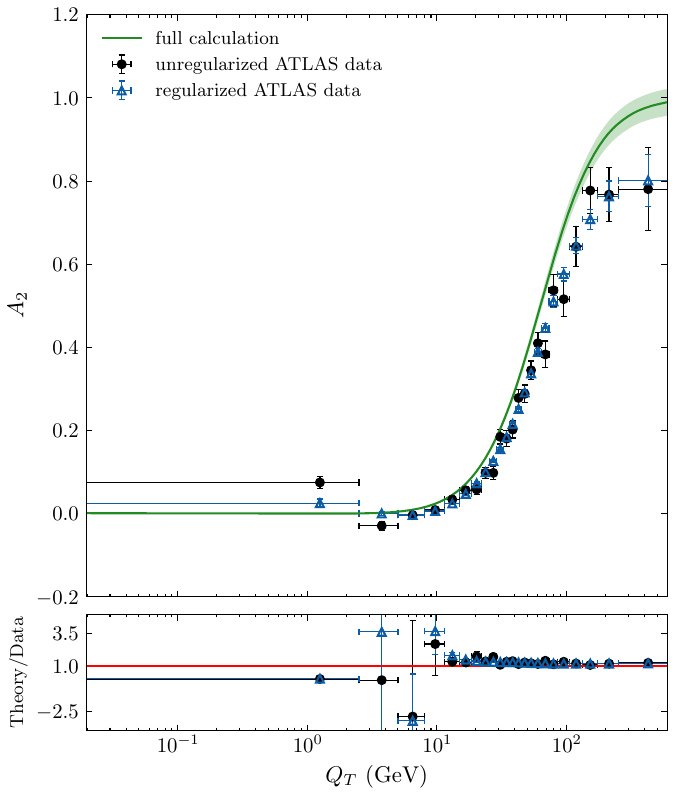}
	\caption{Results for the angular coefficients
	  $A_0$, $A_1$, and $A_2$
          integrated in the specific region of the rapidity $y$
          for different values of $Q_T$ and at $\sqrt{s}=8$ TeV: 
          (a) results for the $A_0$ at $|y| \in [0,3.5]$ (left panel),
          (b) results for the $A_1$ at $|y| \in [0,2]$ (central panel),
          (c) results for the $A_2$ at $|y| \in [0,3.5]$ (right panel).
          Dots and triangles display unregularized and regularized data of
          the ATLAS Collaboration~\cite{ATLAS:2016rnf}, respectively. 
          Our results are indicated by the green shaded bands.}
	\label{A012}	
\end{figure}

\begin{figure}[t]        
        \includegraphics[height=6.5cm]{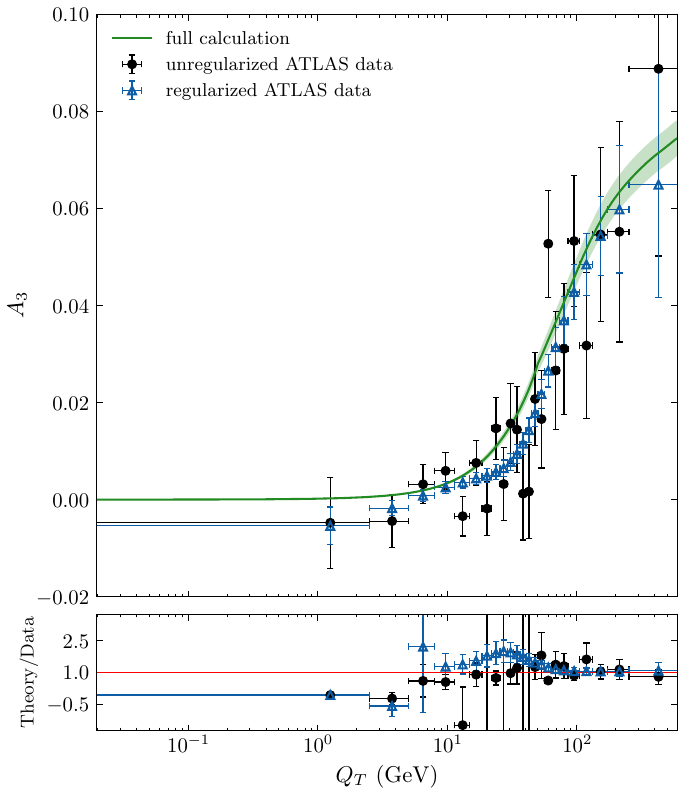}
	\qqqquad
	\includegraphics[height=6.5cm]{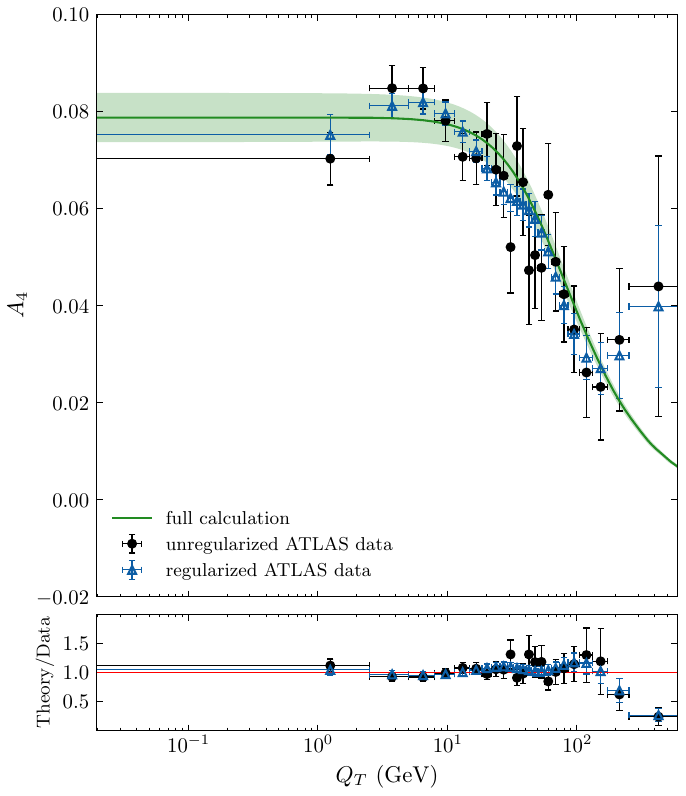}
	\caption{Results for the angular coefficients 
	  $A_3$ and $A_4$ integrated in the specific region
          of the rapidity $y$
          for different values of $Q_T$ and at $\sqrt{s}=8$ TeV:
          (a) results for the $A_3$ at $|y| \in [0,3.5]$ (left panel),
          (b) results for the $A_4$ at $|y| \in [0,3.5]$ (right panel).
          Dots and triangles display unregularized and regularized data of
          the ATLAS Collaboration~\cite{ATLAS:2016rnf}, respectively. 
          Our results are indicated by the green shaded bands.}
   	  \label{A34}	
\end{figure}

We obtained the results for the angular coefficients by direct calculation
Eq.~(\ref{factorization}) for every helicity hadronic structures
with taking into account quark-antiquark and (anti)quark-gluon contribution
at the LO accuracy.
For our purposes we used LHAPDF library~\cite{Buckley:2014ana},
in particular, the CT18NLO~\cite{Hou:2019efy} parametrization for PDFs
including the scale evolution $Q \in [80,100]$ GeV.
We performed a numerical simulation of data
by random selection of normal distribution in the same region of $Q$
as in the ATLAS experiment and for every value of $Q_T$. 
This specifies the uncertainty range of our theoretical prediction
for helicity structure functions and angular coefficients,
which are presented in Figs.~\ref{A012} and~\ref{A34}.
For $T$-even angular coefficients these two sets have similar behavior.
Our predictions are in good agreement with data
(see Figs.~\ref{A012} and~\ref{A34}).

In the LO of the $\alpha_s$ expansion, the LT relation is not violated
and we see that data for the $A_2$ angular coefficient lie below
a theoretical curve (see Fig.~\ref{A012}).
Including NLO $\alpha_s^2$ corrections to the hadronic
structure function we should be able to produce a violation
of the LT relation and it is clearly shown
in the ATLAS paper~\cite{ATLAS:2016rnf}
by using DYNNLO package~\cite{Catani:2009sm}.
Besides, the same analysis is presented
in Ref.~\cite{Gauld:2017tww}. Our analysis of the angular structure
of the DY process at the $\alpha_s^2$ order 
is in progress and will be completed in near future.
As shown in Refs.~\cite{ATLAS:2016rnf,Gauld:2017tww},
taking into account of the $\alpha_s^2$ 
corrections should give sizable contribution to
the $A_2$ angular coefficient. On the other hand, 
it should make a tiny setting of $A_1$, $A_3$ and $A_4$ angular coefficients
by changing of hard part of scattering amplitude and weak coupling.  
	
Results for the coefficient $A_1$, which is related to the single-spin-flip 
hadronic helicity structure function $W_\Delta$, should be corrected and
improved by taking into account of the $\alpha_s^2$ contributions.
The growth of the $A_1$ at large $Q_T > Q$ should be studied directly 
by taking into account a possibility of fragmentation of quarks into
virtual gauge bosons~\cite{Kang:2008wv,Qiu:2001ac,Qiu:2001nr}.  
The $Q_T$ behavior of the $A_3$ and $A_4$ angular coefficients have good 
agreement with data. Values of these coefficients are suppressed due
to smallness of the weak coupling constant. At small $Q_T$, we can present 
a solution regarding relations involving transverse
helicity structure functions $(2-A_0)/2=(G_1/G_2) A_4$ 
and single spin-flip structure functions $A_1=(G_1/G_2) A_3$.
With growth of $Q_T$ the coefficients $A_1$ and $A_3$
are increased due to the factor $\sqrt{1+\rho^2}$. 

\begin{figure}[t]
	\includegraphics[height=6.5cm]{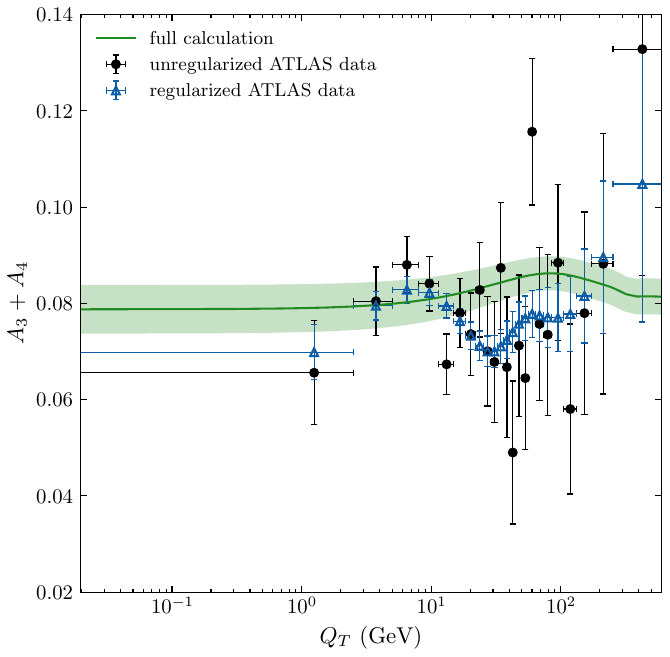}
	\qqqquad
	\includegraphics[height=6.5cm]{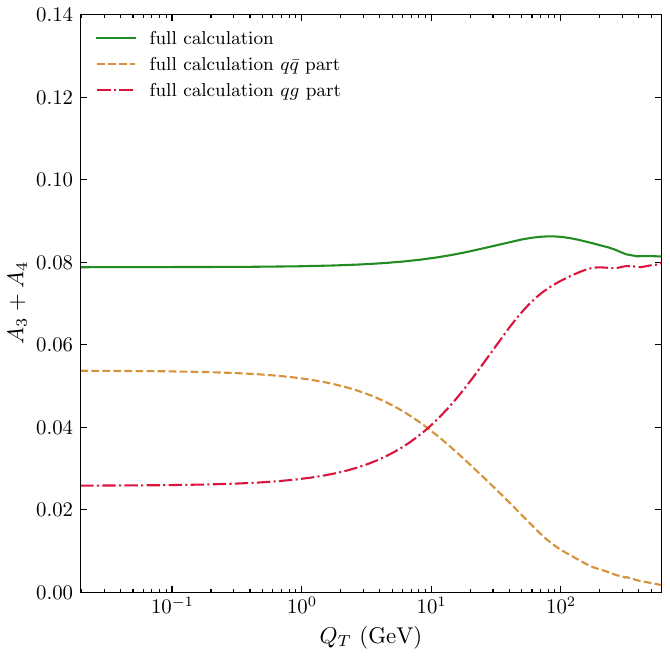}
	\caption{Results for the combination of the angular coefficients 
	  $A_3+A_4$ integrated in the region of the rapidity $y\in [0,3.5]$
          for different values of $Q_T$, for $Q \in [80,100]$ GeV
          and at $\sqrt{s}=8$ TeV:
          (a) total results (left panel),
          (b) central values of the total and partial quark-antiquark
          and quark-gluon contributions
          to the combination $A_3+A_4$ (right panel).
          Dots and triangles display unregularized and regularized data of
          the ATLAS Collaboration~\cite{ATLAS:2016rnf}, respectively. 
          Our results are indicated by the green shaded band.}
	\label{A3plusA4}	
\end{figure}

Herewith, we want to note that the behavior of the combination $A_3+A_4$
in the range of $Q_T$ up to $100$ GeV is stable nearly $G_2/G_1$.
We present the behavior for the combination $A_3+A_4$
in Fig.~\ref{A3plusA4}, we can see that the contribution of
the quark-antiquark subprocess decreases for this combination.
It is connected to a decreasing of the $A_4$ with a growth of $Q_T$.
From the other side, the quark-gluon contribution of $A_3$ angular
coefficient is increasing with a growth of $Q_T$.
Such combinations as $A_3+A_4$ can be also used for analysis
of experimental results.

The angular coefficient $A_4$ is related to the FB asymmetry
$A_4 = \frac{8}{3} \, A_{\rm FB}$ which is important for fixing of
the weak coupling. As it was stressed in Refs.~\cite{Bodek:2010qg,Ball:2022qtp},
the center mass frame for the partonic level can be defined 
only for cases, where we have zero transverse momentum of the lepton pair.
For nonzero values of the leptonic pair transverse momentum,
the partonic level is approximated by
the Collins-Soper frame~\cite{Collins:1977iv}.
Besides, the measurements are needed to recalculate FB asymmetry
for the $pp$ collision.
This is connected with the fact that the quark is defined to be
the direction of the hadron in the DY process. Direction of
antiquarks are needed to be averaged. To simplify extraction,
we need to include a weight factor, which connects rotation
of lepton direction regarding hadron collision frame~\cite{Bodek:2010qg}.
If angle $\theta=0$, then we obtain that $A_{\rm FB}=3/4 \, A_4$.
In the collinear factorization picture, we propose that all partons
have the same direction as hadrons. 
Because of this, we can make a calculation in specific system
where $\theta=0$, which will be the case
similar to the $e^+e^-$ annihilation into hadrons. 

\begin{figure}[t]
	\includegraphics[height=6.5cm]{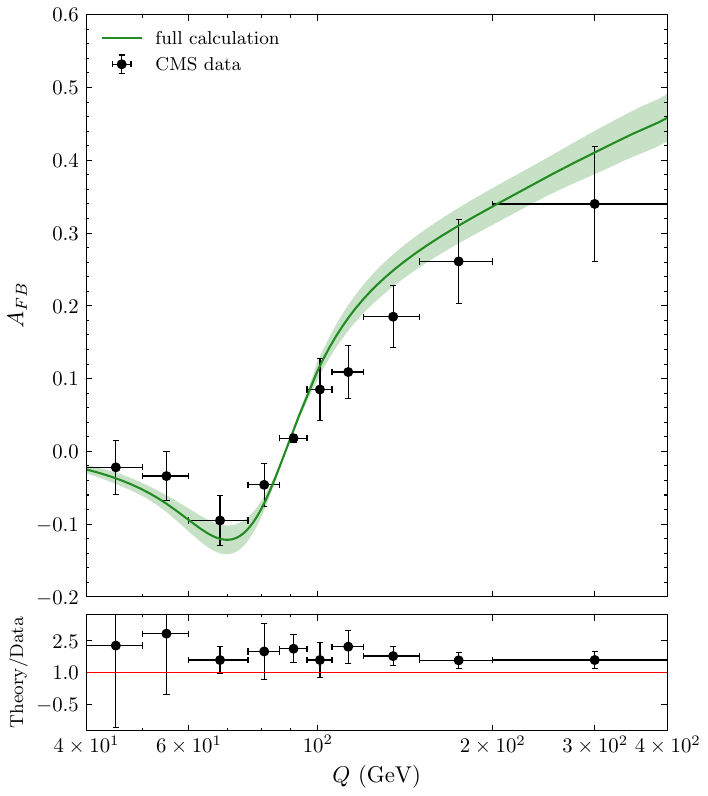}
	\includegraphics[height=6.5cm]{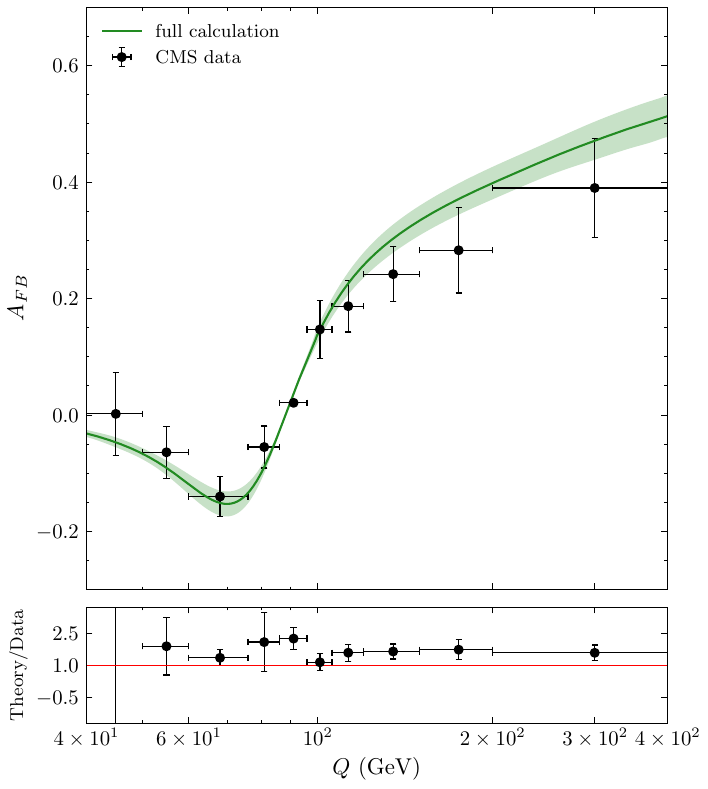}   
	\includegraphics[height=6.5cm]{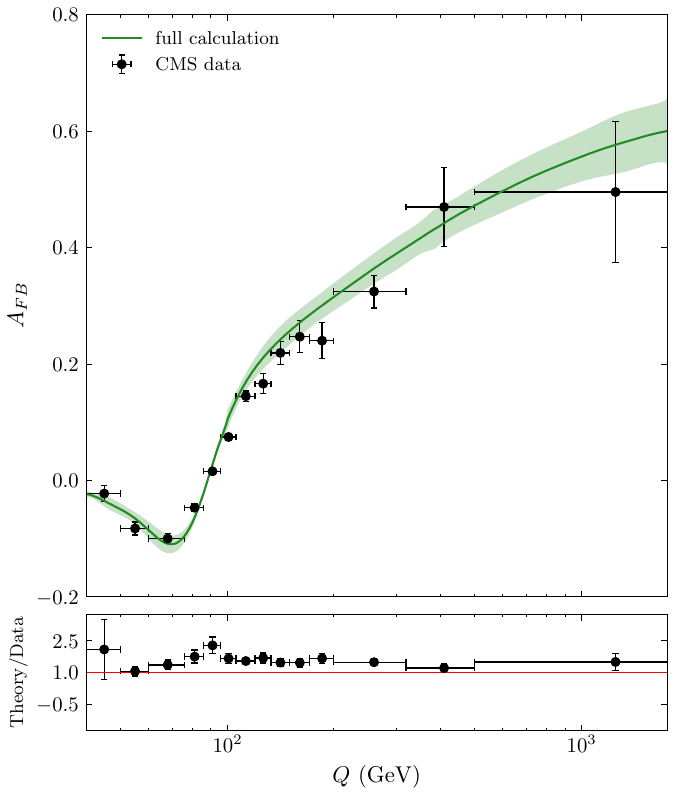}
	\caption{Results for the FB asymmetry $A_{\rm FB}$ as
                 function of $Q$ in comparison
                 with data extracted by  
		 the CMS Collaboration:
                 (a) for rapidity ranges $1 < |y| < 1.25$
                 and at $\sqrt{s}=7$ TeV~\cite{CMS:2012zgh}
                 (left panel),
                 (b) for rapidity ranges $1.25 < |y| < 1.5$
                 and at $\sqrt{s}=7$ TeV~\cite{CMS:2012zgh}
                 (central panel),
		 (c) for rapidity ranges $1 < |y| < 1.25$ and
                 at  $\sqrt{s}=8$ TeV~\cite{CMS:2016bil} (right panel).} 
	         \label{Afb_fig}	
\end{figure}

We show results for the FB asymmetry in Fig.~\ref{Afb_fig},
where we present the behavior of the $A_{\rm FB}=\frac{3}{4} \, A_4$
at different invariant masses of lepton pair. Experimental points
correspond to data obtained by the CMS Collaboration
at $\sqrt{s}=7$ TeV~\cite{CMS:2012zgh}
and $\sqrt{s}=8$ TeV~\cite{CMS:2016bil} for rapidity
ranges $1<|y|<1.25$, $1.25<|y|<1.5$.
We also take into account that data were obtained
for $Q_T > 20$ GeV. Upper limit for $Q_T$ in our numerical analysis
is 100 GeV. 

 \begin{figure}[t]
	\centering
	\begin{minipage}{.45\linewidth}
	\includegraphics[height=6.5cm]{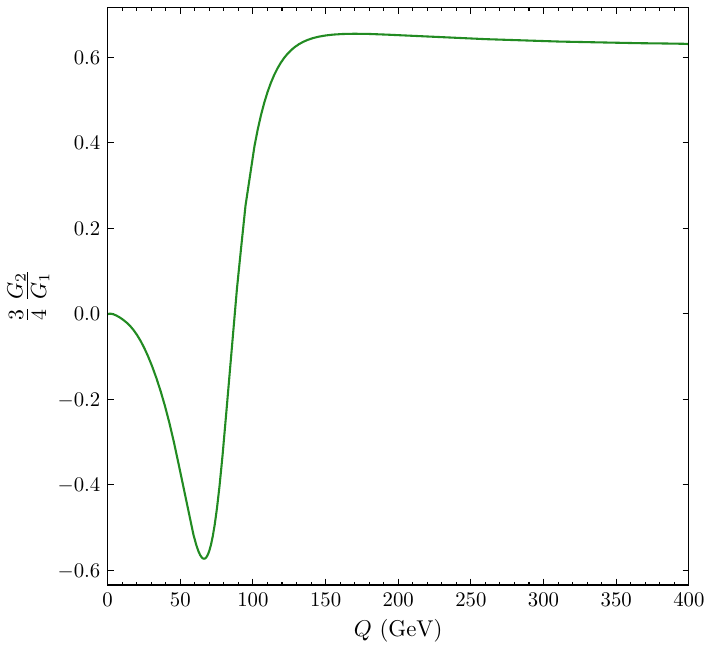}
\caption{The $\frac{3}{4} \, \frac{G_2}{G_1}$
	as function of $Q$.}
\label{G2G1}
	\end{minipage}
	\hspace{.05\linewidth}
	\begin{minipage}{.45\linewidth}
	       \includegraphics[height=6.5cm]{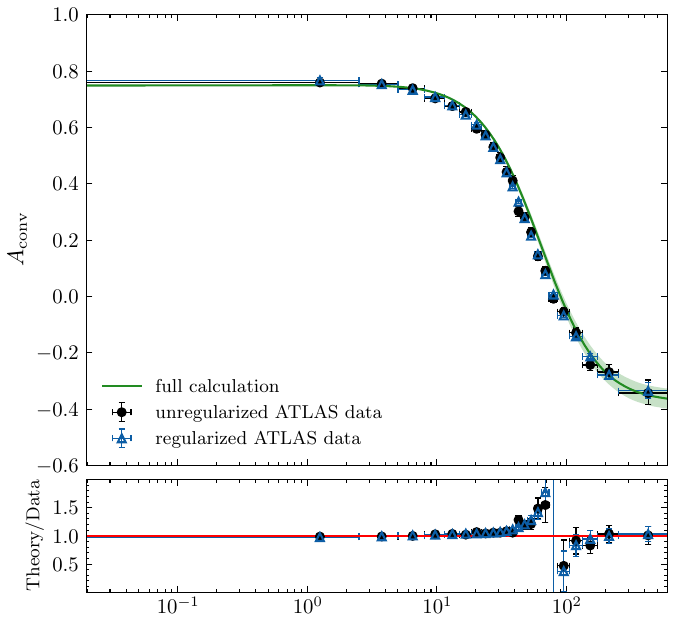}
	\caption{Results for the convexity $A_{\rm conv}$
		for rapidity range $0 < |y| < 3.5$
		as function of $Q_T$.}
	\label{Aconv}	
	\end{minipage}
 \end{figure}
 
We see that the FB asymmetry describes a behavior of the ratio of
the couplings $\frac{G_2}{G_1}$
with an additional factor $\frac{3}{4}$ in front. 
We show the $Q$ dependence of the $\frac{3}{4} \, \frac{G_2}{G_1}$
ratio on Fig.~\ref{G2G1}.
At large $Q$ and small $Q_T$, the helicity structure function
$W_L \sim O(\rho^2)$ will be suppressed in comparison
with $W_T$ and $W_{T_P}$ hadronic structure functions,
which have additional factor $1/\rho^2=Q^2/Q_T^2$.
In this region, the $W_T$ and $W_{T_P}$
hadronic structure functions will be equal (see equations
in Sec.~\ref{small_Qt}) and $A_{\rm FB}$ will approach to the
limit $\frac{3}{4} \, \frac{G_2}{G_1}$. Finally,
in Fig.~\ref{Aconv} for $\sqrt{s} = 8$ TeV and at $|y| \in [0,3.5]$,
we show our predictions for the convexity $A_{\rm conv}$,
which encodes the $W_T-W_L$ asymmetry of the transverse and
longitudinal hadronic structure functions.
One can see that for given values of kinematical parameters 
the $A_{\rm conv}$ crosses zero at $Q_T \simeq 90$ GeV, which
corresponds to the critical point, where $W_T = W_L$, $A_0 = 2/3$, and
$\lambda = 0$.


\section{Conclusion}
\label{Conclusion}

We presented analytical results for the Drell-Yan $T$-even hadronic structure
functions in the framework of the pQCD based on the collinear factorization
scheme and at the leading order in the $\alpha_s$ expansion.
We obtained exact and full analytical formula for the small $Q_T$ expansion of
the hadronic structure functions without referring to the specific order of
such expansion. We also show how our formalism can be extended to the study
of other QCD processes, e.g., such as the SIDIS process. 

We demonstrated that our full results in leading order in the $\alpha_s$
expansion are in good agreement with presently available data from
the ATLAS Collaboration~\cite{ATLAS:2016rnf} for the angular coefficients
in the DY process at $\sqrt{s}=8$ TeV.
Additionally we presented analysis for the FB asymmetry and comparison
with data. We pointed out that the small $Q_T/Q$ limit plays an important
role for the FB asymmetry. In near future we plan to study full rapidity
dependence of the angular coefficients occurring in the DY process and
extend our analysis to the $\alpha_s^2$ order in strong coupling expansion. 

\begin{acknowledgments} 

A.Z. thanks T\"{u}bingen University for warm hospitality
during his visits in the period which lasted from 2017-2024, 
when a part of the work was done and
supported by the DAAD scholarships No. 57314021 and No. 57693448. 
This work was funded by ANID$-$Millen\-nium Program$-$ICN2019\_044 (Chile).
We thank Werner Vogelsang, J\"{u}rgen K\"{o}rner,
Sergey Kuleshov, Fabian Wunder, Vladimir Saleev, Andrey Arbuzov,
and Arie Bodek for discussions and interest to our work. 
	
\end{acknowledgments}

\appendix
\section{Kinematics} 
\label{app:Kinematics}

Here we specify kinematics of the DY process on both 
hadronic and partonic level. Hadronic center of mass (CM) frame 
is specified by the following 
choice of hadronic $P_1,P_2$ and the finale vector boson $q$ momenta 
\eq 
& &
P_1^\mu = \sqrt{\frac{s}{2}} \, n_+^\mu\,, \quad 
P_2^\mu = \sqrt{\frac{s}{2}} \, n_-^\mu\,, \nonumber\\ 
& & 
q^\mu = Q^+ n_+^\mu + Q^- n_-^\mu + Q_T n_T^\mu \,, 
\en 
where $n_i^\mu$ are unit vectors specifying the light-cone 
coordinates in CM frame 
\eq 
& &n_\pm^\mu = \delta^{\mu\pm}\,, \quad 
n_T^\mu = \delta^{\mu T}\,, \nonumber \\
& &n^2_\pm = 0\,, \quad n^2_T=-1\,, \quad 
n_+ \cdot n_- = 1\,, \quad n_\pm \cdot n_T = 0 ,  
\en 
and 
\eq 
Q^\pm =  x_{1,2} \sqrt{\frac{s}{2}} = e^{\pm y} \sqrt{\frac{Q^2+Q_T^2}{2}},
\en
where $y$ is the rapidity 
\eq 
y = \frac{1}{2} \, \log\frac{x_1}{x_2} = \frac{1}{2} \, \log\frac{q^0+q^3}{q^0-q^3} \,. 
\en 

Current conserving Minkowski tensor $\tilde g_{\mu\nu}$ 
and hadronic momenta $\tilde P_{1,2}^\mu$ are defined as 
\eq 
& &\tilde g_{\mu\nu} = g_{\mu\nu} - \frac{q_\mu q_\nu}{q^2}\,, 
\quad q^\mu \tilde g_{\mu\nu} = 0 \,, \nonumber\\
& &\tilde P_{i\mu} = \tilde g_{\mu\nu} \frac{P_i^\nu}{\sqrt{s}} \,, 
\quad q^\mu \tilde P_{i\mu} = 0 .
\en 

Next we introduce the set of invariant variables 
independent on the frame 
\eq 
q_{P_1}  &=&  \frac{P_1q}{\sqrt{s}} = 
x_2 \frac{\sqrt{s}}{2}\,, 
\quad 
q_{P_2} \ = \ \frac{P_2q}{\sqrt{s}} = 
x_1 \frac{\sqrt{s}}{2}\,, 
\nonumber\\
q_P &=& q_{P_1} + q_{P_2} = \frac{Pq}{\sqrt{s}} 
= (x_1+x_2) \frac{\sqrt{s}}{2}  
\,, \\
q_p &=& - q_{P_1} + q_{P_2} = \frac{pq}{\sqrt{s}} 
= (x_1-x_2) \frac{\sqrt{s}}{2}  
\,, \nonumber
\en
where $P = P_1 + P_2$ and $p = - P_1 + P_2$. 
 
Hadron-level Mandelstam variables: 
\eq 
s &=& (P_1 + P_2)^2\,,\nonumber\\
t &=& (P_1 - q)^2 = P_1^2 + Q^2 - 2 P_1 q = 
Q^2 - x_2 s 
\,, \nonumber\\
u &=& (P_2 - q)^2 = P_2^2 + Q^2 - 2 P_2 q = 
Q^2 - x_1 s 
\,, \nonumber\\
s + t + u &=& s + 2 Q^2 - 2 P q =  
s (1 - x_1 - x_2) + 2 Q^2 \,. 
\en 
Parton-level Mandelstam variables: 
\eq 
\hat{s} &=& (p_1 + p_2)^2 = \xi_1 \xi_2 s \,, \nonumber\\
\hat{t} &=& (p_1 - q)^2 = 
Q^2 - \xi_1 x_2 s  
= Q^2 - \xi_1 (Q^2 - t) \,, \nonumber\\
\hat{u} &=& (p_2 - q)^2 = 
Q^2 - \xi_2 x_1 s  
= Q^2 - \xi_2 (Q^2 - u) \,, \nonumber\\
\hat{s} + \hat{t} + \hat{u} &=& \xi_1 \xi_2 s + \xi_1 t + \xi_2 u 
+ Q^2 (2 - \xi_1 - \xi_2) \nonumber\\
&=& s (\xi_1 - x_1) (\xi_2 - x_2) - s x_1 x_2 + 2 Q^2  = Q^2 \,,
\en
where in hadronic CM frame we have 
\eq 
Q^2   &=& x_1 x_2 s - Q_T^2 \,, \nonumber\\
Q_T^2 &=& s (\xi_1 - x_1) (\xi_2 - x_2) \,. 
\en 
Using the fraction parameters $z_i = x_i/\xi_i$ one gets
\eq 
\hat{s} &=& \frac{Q^2 + Q_T^2}{z_1 z_2} = 
\frac{Q_T^2}{(1-z_1)(1-z_2)} \,,\nonumber\\
\hat{t} &=& Q^2 - \frac{Q^2 + Q_T^2}{z_1} 
= - \frac{Q_T^2}{1-z_2} \,, \nonumber\\
\hat{u} &=& Q^2 - \frac{Q^2 + Q_T^2}{z_2} 
= - \frac{Q_T^2}{1-z_1} \,, \nonumber\\
\hat{s} + \hat{t} + \hat{u} &=& Q^2 = 
Q_T^2 \frac{z_1+z_2-1}{(1-z_1)(1-z_2)} \,.
\en

\section{Helicity hadronic and leptonic structure functions}
\label{app:Wstr}

Covariant hadronic $W_{\mu\nu}$ and leptonic
$L_{\mu\nu}$ and leptonic tensors can be related
to the corresponding helicity tensors
$W_{\lambda\lambda^\prime}$ and
$L_{\lambda\lambda^\prime}$ and~\cite{Lam:1978pu,Collins:1977iv,%
Mirkes:1992hu,Boer:2006eq,Berger:2007jw,Lyubovitskij:2024civ} 
with the use of the gauge boson polarization vectors	
$\epsilon_\lambda^\mu(q)$:
 \eq 
 F_{\lambda\lambda^\prime} = \epsilon^{\dagger\mu}_\lambda \, 
 F_{\mu\nu} \, \epsilon^{\nu}_{\lambda^\prime} \,, \quad F = W, L\,.
 \en
 Both covariant tensors have the similar expansion
 in terms of helicity tensors: 
 \eq 
 W^{\mu\nu} &=& 
 W_T \Big( 
 \epsilon^{\mu}_+ \epsilon^{\ast\nu}_+ 
 + \epsilon^{\mu}_- \epsilon^{\ast\nu}_- 
 \Big) 
 \, + \, W_{T_P} \Big( 
 \epsilon^{\mu}_+ \epsilon^{\ast\nu}_+
 - \epsilon^{\mu}_- \epsilon^{\ast\nu}_- 
 \Big) 
 \, + \, 
 W_L \epsilon^{\mu}_0 \epsilon^{\ast\nu}_0  
 \nonumber\\[4mm]
 &+&
 W_{\Delta\Delta} \Big( 
 \epsilon^{\mu}_+ \epsilon^{\ast\nu}_- 
 + \epsilon^{\mu}_- \epsilon^{\ast\nu}_+ 
 \Big) 
 \, + \, i W_{\Delta\Delta_P} 
 \Big( 
 \epsilon^{\mu}_+ \epsilon^{\ast\nu}_-
 - \epsilon^{\mu}_- \epsilon^{\ast\nu}_+
 \Big) \nonumber\\[4mm]
 &+&
 W_{\Delta} \Big( 
 \frac{\epsilon^{\mu}_+ - \epsilon^{\mu}_-}{\sqrt{2}} 
 \,  \epsilon^{\ast\nu}_0 \, + \, 
 \epsilon^{\mu}_0 \,    
 \frac{\epsilon^{\ast\nu}_+ - \epsilon^{\ast\nu}_-}{\sqrt{2}} \Big) 
 \, + \, 
 i W_{\Delta_P}  
 \Big( 
 \frac{\epsilon^{\mu}_+ + \epsilon^{\mu}_-}{\sqrt{2}} 
 \,  \epsilon^{\ast\nu}_0 \, - \, 
 \epsilon^{\mu}_0 \,    
 \frac{\epsilon^{\ast\nu}_+ + \epsilon^{\ast\nu}_-}{\sqrt{2}} \Big) 
 \nonumber\\[4mm]
 &+&
 i W_{\nabla} 
 \Big( 
 \frac{\epsilon^{\mu}_+ - \epsilon^{\mu}_-}{\sqrt{2}} 
 \,  \epsilon^{\ast\nu}_0 \, - \, 
 \epsilon^{\mu}_0 \,    
 \frac{\epsilon^{\ast\nu}_+ - \epsilon^{\ast\nu}_-}{\sqrt{2}} \Big) 
 \, + \, 
 W_{\nabla_P} 
 \Big( 
 \frac{\epsilon^{\mu}_+ + \epsilon^{\mu}_-}{\sqrt{2}} 
 \,  \epsilon^{\ast\nu}_0 \, + \, 
 \epsilon^{\mu}_0 \,    
 \frac{\epsilon^{\ast\nu}_+ + \epsilon^{\ast\nu}_-}{\sqrt{2}} \Big) \, ,
 \label{helicity_W}
 \en
 and
 \eq 
 L^{\mu\nu} &=& 
 L_T \Big( 
 \epsilon^{\mu}_+ \epsilon^{\ast\nu}_+ 
 + \epsilon^{\mu}_- \epsilon^{\ast\nu}_- 
 \Big) 
 \, + \, L_{T_P} \Big( 
 \epsilon^{\mu}_+ \epsilon^{\ast\nu}_+
 - \epsilon^{\mu}_- \epsilon^{\ast\nu}_- 
 \Big) 
 \, + \, 
 L_L \epsilon^{\mu}_0 \epsilon^{\ast\nu}_0  
 \nonumber\\[4mm]
 &+&
 L_{\Delta\Delta} \Big( 
 \epsilon^{\mu}_+ \epsilon^{\ast\nu}_- 
 + \epsilon^{\mu}_- \epsilon^{\ast\nu}_+ 
 \Big) 
 \, + \, i L_{\Delta\Delta_P} 
 \Big( 
 \epsilon^{\mu}_+ \epsilon^{\ast\nu}_-
 - \epsilon^{\mu}_- \epsilon^{\ast\nu}_+
 \Big) \nonumber\\[4mm]
 &+&
 L_{\Delta} \Big( 
 \frac{\epsilon^{\mu}_+ - \epsilon^{\mu}_-}{\sqrt{2}} 
 \,  \epsilon^{\ast\nu}_0 \, + \, 
 \epsilon^{\mu}_0 \,    
 \frac{\epsilon^{\ast\nu}_+ - \epsilon^{\ast\nu}_-}{\sqrt{2}} \Big) 
 \, + \, 
 i L_{\Delta_P}  
 \Big( 
 \frac{\epsilon^{\mu}_+ + \epsilon^{\mu}_-}{\sqrt{2}} 
 \,  \epsilon^{\ast\nu}_0 \, - \, 
 \epsilon^{\mu}_0 \,    
 \frac{\epsilon^{\ast\nu}_+ + \epsilon^{\ast\nu}_-}{\sqrt{2}} \Big) 
 \nonumber\\[4mm]
 &+&
 i L_{\nabla} 
 \Big( 
 \frac{\epsilon^{\mu}_+ - \epsilon^{\mu}_-}{\sqrt{2}} 
 \,  \epsilon^{\ast\nu}_0 \, - \, 
 \epsilon^{\mu}_0 \,    
 \frac{\epsilon^{\ast\nu}_+ - \epsilon^{\ast\nu}_-}{\sqrt{2}} \Big) 
 \, + \, 
 L_{\nabla_P} 
 \Big( 
 \frac{\epsilon^{\mu}_+ + \epsilon^{\mu}_-}{\sqrt{2}} 
 \,  \epsilon^{\ast\nu}_0 \, + \, 
 \epsilon^{\mu}_0 \,    
 \frac{\epsilon^{\ast\nu}_+ + \epsilon^{\ast\nu}_-}{\sqrt{2}} \Big) \,, 
 \label{helicity_L}
 \en
where lepton helicity structure functions are defined as
\eq\label{LI_str_fun}
L_T &=& L_{++} + L_{--} = 
L_{\mu\nu} \, (X^\mu X^\nu +  Y^\mu Y^\nu) =Q^2 (1 + \cos^2\theta) 
\,, \nonumber\\[4mm]
L_L &=& L_{00} = L_{\mu\nu} \, Z^\mu Z^\nu = Q^2 (1 - \cos^2\theta) \,,
\nonumber\\[4mm]
L_{\Delta\Delta} &=& i \, (L_{+-} - L_{-+}) = 
L_{\mu\nu} \, (Y^\mu Y^\nu -  X^\mu X^\nu) 
= Q^2 \sin^2\theta\cos 2\phi 
\,, \nonumber\\ [4mm]
L_{\Delta} &=& \frac{1}{\sqrt{2}} \Big(L_{+0} - L_{-0} + L_{0+} - L_{0-}\Big)  
= - L_{\mu\nu} \, \Big( X^\mu Z^\nu + Z^\mu X^\nu \Big) 
= Q^2 \sin 2\theta \, \cos\phi 
\,, \nonumber\\[4mm]
L_{\nabla} &=& \frac{i}{\sqrt{2}} \Big(L_{+0} - L_{-0} - L_{0+} + L_{0-}\Big)
= i  \, L_{\mu\nu} \, \Big( Z^\mu X^\nu - X^\mu Z^\nu \Big) 
= Q^2 \sin\theta \, \sin\phi 
\,, \nonumber\\[4mm] 
L_{T_P} &=& L_{++} - L_{--}
=  i \, L_{\mu\nu} \, \Big( X^\mu Y^\nu - Y^\mu X^\nu \Big) = Q^2 \cos\theta 
\,, \nonumber\\[4mm]
L_{\Delta\Delta_P} &=&
- L_{\mu\nu} \, \Big( X^\mu Y^\nu + Y^\mu X^\nu \Big) = Q^2 \sin^2\theta \, \sin 2\phi 
\,, \nonumber\\ 
L_{\Delta_P} &=& \frac{i}{\sqrt{2}} \Big(L_{+0} + L_{-0} - L_{0+} - L_{0-}\Big) 
= - L_{\mu\nu} \, \Big( Y^\mu Z^\nu + Z^\mu Y^\nu \Big) 
= Q^2 \sin 2\theta \, \sin\phi 
\,, \nonumber\\[4mm]
L_{\nabla_P} &=& \frac{1}{\sqrt{2}} \Big(L_{+0} + L_{-0} + L_{0+} + L_{0-}\Big) 
= i  \, L_{\mu\nu} \, \Big( Y^\mu Z^\nu - Z^\mu Y^\nu \Big) 
= Q^2 \sin\theta \, \cos\phi \,.
\en 
 
\section{Relations between different sets of the structure functions}
\label{app:Str_Func}

Three sets of the structure functions $\{A_i\}$, $\{W_i\}$ and 
$\{\lambda,\mu,\nu,\ldots\}$ are related 
as~\cite{Lam:1978pu,Collins:1977iv,Mirkes:1992hu,%
Boer:2006eq,Berger:2007jw,Lyubovitskij:2024civ}  
\eq 
\lambda &=& \frac{W_T-W_L}{W_T+W_L} = \frac{2-3A_0}{2+A_0}\,, 
\quad 
\mu \, = \, \frac{W_\Delta}{W_T+W_L} = \frac{2A_1}{2+A_0}\,, 
\quad 
\nu \, = \, \frac{2W_{\Delta\Delta}}{W_T+W_L} = \frac{2A_2}{2+A_0}\,, 
\nonumber\\[4mm] 
\tau &=& \frac{W_{\nabla_P}}{W_T+W_L} = \frac{2A_3}{2+A_0}\,, 
\quad \hspace*{.2cm}
\eta \, = \, \frac{W_{T_P}}{W_T+W_L} = \frac{2A_4}{2+A_0}\,, 
\quad \hspace*{.1cm}
\xi \, = \, \frac{W_{\Delta\Delta_P}}{W_T+W_L} = \frac{2A_5}{2+A_0}\,, 
\nonumber\\[4mm] 
\zeta &=& \frac{W_{\Delta_P}}{W_T+W_L} = \frac{2A_6}{2+A_0}\,, 
\quad \hspace*{.15cm}
\chi \, = \, \frac{W_{\nabla}}{W_T+W_L} = \frac{2A_7}{2+A_0} 
\en 
or 
\eq 
A_0 &=& \frac{2W_L}{2W_T+W_L} = \frac{2 (1-\lambda)}{3+\lambda} \,, 
\quad 
A_1 \, = \, \frac{2W_\Delta}{2W_T+W_L} = \frac{4 \mu}{3+\lambda} \,, 
\quad 
A_2 \, = \, \frac{4W_{\Delta\Delta}}{2W_T+W_L} = \frac{4 \nu}{3+\lambda} \,, 
\nonumber\\[4mm] 
A_3 &=& \frac{2W_{\nabla_P}}{2W_T+W_L} = \frac{4 \tau}{3+\lambda} \,, 
\quad \hspace*{.4cm}
A_4 \, = \, 
\frac{2W_{T_P}}{2W_T+W_L} = \frac{4 \eta}{3+\lambda} \,, 
\quad \hspace*{.05cm}
A_5 \, = \,  \frac{2W_{\Delta\Delta_P}}{2W_T+W_L} = \frac{4 \xi}{3+\lambda} \,, 
\nonumber\\[4mm] 
A_6 &=& \frac{2W_{\Delta_P}}{2W_T+W_L} = \frac{4 \zeta}{3+\lambda} \,, 
\quad \hspace*{.4cm}
A_7 \, = \, \frac{2W_{\nabla}}{2W_T+W_L} = \frac{4 \chi}{3+\lambda} \,. 
\en 

\section{Small $Q_T$ expansion of the helicity hadronic structure functions}
\label{app:Expansion}

In this Appendix we present some additional details and results for the $Q_T$
expansion of the helicity hadronic structure functions.

In particular, we present the complete and universal formula for the
partial $x$ derivatives of the hadronic/partonic structres functions.
As we stressed before, the main task here is to make the partial
derivative of desired order acting on the convolution
of the perturbative coefficient function and PDF.
Such perturbative function could contain the product
of the regular function $\tau(z)$ and 
possible singularities due to logarithms
$\log^k(1-z)$ and $1/(1-z)^m$ poles.
The regular function $\tau(z)$ can be expanded in the
Taylor series around $z=1$ in order to reduce
the perturbative function to the sum of the terms
containing only log-terms and distributions:
\eq
\biggl[\frac{\log^k(1-z)}{(1-z)^{m}}\biggr]_{+,m-1}
\, \tau(z) &=&
\biggl[\frac{\log^k(1-z)}{(1-z)^{m}}\biggr]_{+,m-1}
\ \sum\limits_{s=0}^{\ell} \,
\frac{1}{s!} \, (z-1)^s \, \partial^{s}_z \tau(1) 
\nonumber\\
&=& \sum\limits_{s=0}^{\ell} \, \frac{(-1)^s}{s!}   
\ \partial^{s}_z \tau(1)
\  \biggl[\frac{\log^k(1-z)}{(1-z)^{m-l}}\biggr]_{+,m-l-1}
\,. 
\en 
Therefore, our task is reduced to
calculation of the following generic integral over $z$
\eq
I(k,m) = \int\limits_x^1 \frac{dz}{z} \,
\biggl[\frac{\log^k(1-z)}{(1-z)^{m}}\biggr]_{+,m-1}
\, f(x/z) \,,
\en
where $f(x/z)$ is the PDF. Then, the master
formula for the nth partial derivative of the
integral $I(k,m)$ and for $n \le k$ reads 
\eq
\frac{\partial^n I(k,m)}{\partial x^n} =
\frac{1}{x^n} \,  \int\limits_x^1 dz \,
\biggl[\frac{D_1(k,m,n;z)}{(1-z)^{m+n}}\biggr]_{+,m+n-1}
\, \Big[f(x/z) \, z^{n-1}\Big]\,,
\en 
where
\eq
D_1(k,m,n;z) = \frac{(m+n-1)!}{(m-1)!} 
\, \sum\limits_{i=0}^n  \, (-1)^i 
\, \frac{k!}{(k-i)!} 
\, \log^{k-i}(1-z) \ T_i 
\en 
and
\eq
T_i = \left\{
\begin{array}{cl} 
1\,, & \quad i = 0 \\[3mm]
\sum\limits_{k_r > \ldots > k_1 = 0}^{n-1}  \,
\prod\limits_{j=1}^i  \, \frac{1}{m+k_j}\,,
& \quad i \ge 1 \,.
\end{array}
\right.
\en 
For the $n > k$ the master formula reads
\eq
\frac{\partial^n I(k,m)}{\partial x^n} =
\frac{1}{x^n} \,  \int\limits_x^1 dz \,
\biggl[\frac{D_2(k,m,n;z)}{(1-z)^{m+n}}\biggr]_{+,m+n-1}
\, \Big[f(x/z) \, z^{n-1}\Big] + \Delta(k,m,n)\,,
\en 
where
\eq
D_2(k,m,n;z) &=& \frac{(m+n-1)!}{(m-1)!} 
\, \sum\limits_{i=0}^{k}  \, (-1)^i
\, \frac{k!}{(k-i)!}
\, \log^{k-i}(1-z) \ T_i
\nonumber\\
&=&
\frac{(m+n-1)!}{(m-1)!} \, \biggl(
\, \sum\limits_{i=0}^{k-1}  \, (-1)^i
\, \frac{k!}{(k-i)!}
\, \log^{k-i}(1-z) \ T_i + (-1)^k \ T_k 
\biggr) 
\en 
and
\eq
\Delta(k,m,n) =
\lim\limits_{z \to 1} \, \sum\limits_{l=k+1}^{n} \,
\left(\frac{(-1)^{m+l}}{x^l \, (m-1)!  \, (m+l-1)} \,
\frac{\partial^{m+l-1}}{\partial z^{m+l-1}}
\, \biggl[z^{l-2} \, f\Big(\frac{x}{z}\Big)\biggr]
\right)^{(n-l)}_x
\,.
\en

Now we present the analytical results for the NLP hadronic structure
functions.
In the case of the quark-antiquark annihilation and quark-gluon Compton
scattering subprocesses we get the same
relations between hadronic structure functions as for LP functions: 
\eq 
W^{{\rm NLP}; ab}_{T}(x_1^0,x_2^0,L_\rho) &=&
\frac{g_{ab; 1}}{g_{ab; 2}} \, 
W^{{\rm NLP}; ab}_{T_P}(x_1^0,x_2^0,L_\rho)
\,,\nonumber\\[2mm]
W^{{\rm NLP}; ab}_{L}(x_1^0,x_2^0,L_\rho) &=&
2 W^{{\rm NLP}; ab}_{\Delta\Delta}(x_1^0,x_2^0,L_\rho) 
\,,\nonumber\\[2mm]
W^{{\rm NLP}; ab}_{\Delta}(x_1^0,x_2^0,L_\rho) &=&
\frac{g_{ab; 1}}{g_{ab; 2}} \, 
W^{{\rm NLP}; ab}_{\nabla_P}(x_1^0,x_2^0,L_\rho) \,,
\en
where $ab = q \bar q, q g$. 

Also there interesting relation between $W^{{\rm NLP}; ab}_{T}$ and
$W^{{\rm NLP}; ab}_{L} = 2 W^{{\rm NLP}; ab}_{\Delta\Delta}$ 
functions. In particular, one can express
$W^{{\rm NLP}; ab}_{L} = 2 W^{{\rm NLP}; ab}_{\Delta\Delta}$ through
combination of $W^{{\rm LP}; ab}_{T}$ and  $W^{{\rm NLP}; ab}_{T}$ as 
\eq
W^{{\rm NLP}; q \bar q}_{L}(x_1^0,x_2^0,L_\rho) &=&
\rho^2 \, \biggl(W^{{\rm NLP}; q \bar q}_{T}
- \frac{1}{2} W^{{\rm LP}; q \bar q}_{T}\biggr)
\,,\nonumber\\[4mm]
W^{{\rm NLP}; q g}_{L}(x_1^0,x_2^0,L_\rho) &=&
\rho^2 \, \biggl(W^{{\rm NLP}; q g}_{T}
- \frac{1}{2} W^{{\rm LP}; q g}_{T}\biggr)
+ 4 \, g_{qq; 1} \, \biggl[\Big(- \frac{1}{2} 
+ L_\rho -  L_\rho \, x_1^0 \partial_{x_1^0}\biggr)
\, q(x_1^0) \, g(x_2^0)
\nonumber\\[4mm]
&-& \int\limits_{x_1^0}^1 \frac{dz_1}{(1-z_1)_+}
q\Big(\frac{x_1^0}{z_1}\Big) g(x_2^0)
+ \int\limits_{x_2^0}^1 \frac{dz_2}{(1-z_2)_+} \, 
\Big(1 - \frac{1+z_2}{2} x_2^0 \partial_{x_2^0}\Big)
q(x_1^0) g\Big(\frac{x_2^0}{z_2}\Big)\biggr] \,. 
\en

We remind that the hadronic structure functions at any order of
small $Q_T$ expansion are given by
\eq\label{Wi_power}
W^{{\rm N}^m{\rm LP}}(x_1^0, x_2^0,L_\rho) &=& \frac{1}{x_1^0x_2^0}\sum_{a,b}
\, \biggl[
   R_{ab}^{{\rm N}^m{\rm LP}}(x_1^0,x_2^0,L_\rho)
\, f_{a/H_1}(x_1^0)
\, f_{b/H_2}(x_2^0)
\nonumber\\[4mm]
&+& \Big(P_{ba}^{{\rm N}^m{\rm LP}} \otimes f_{b/H_2}\Big)(x_2^0,x_1^0,L_\rho)
\ f_{a/H_1}(x_1^0)
\nonumber\\[4mm]
&+& \Big(P_{ab}^{{\rm N}^m{\rm LP}} \otimes f_{a/H_1}\Big)(x_1^0,x_2^0,L_\rho)
\ f_{b/H_2}(x_2^0) 
\biggr] 
\,.
\en

It is convenient to expand the perturbative functions $R_{ab}^{{\rm N}^m{\rm LP}}$,
$P_{ab}^{{\rm N}^m{\rm LP}}(z_1,x_2^0,L_\rho)$, and
$P_{ba}^{{\rm N}^m{\rm LP}}(z_2,x_1^0,L_\rho)$ as
\eq\label{RP_exp}
R_{ab}^{{\rm N}^m{\rm LP}}(x_1^0,x_2^0,L_\rho) &=& \sum\limits_{s_1,s_2 = 0}^{m}
\, R_{ab; s_1s_2}^{{\rm N}^m{\rm LP}}(L_\rho) \, T^R_{s_1s_2}(x_1^0,x_2^0) \,,
\nonumber\\[4mm]
P_{ab}^{{\rm N}^m{\rm LP}}(z_1,x_2^0,L_\rho) 
&=& \sum\limits_{s=0}^{m}
\, P_{a, 1s}^{{\rm N}^m{\rm LP}}(z_1,L_\rho) \, T^P_s(x_2^0) \,,
\nonumber\\[4mm]
P_{ba}^{{\rm N}^m{\rm LP}}(z_2,x_1^0,L_\rho) 
&=& \sum\limits_{s=0}^{m}
\, P_{ab,2s}^{{\rm N}^m{\rm LP}}(z_2,L_\rho) \, T^P_s(x_1^0) \,,
\en
where
\eq\label{RP_exp2}
T^R_{s_1s_2}(x_1^0,x_2^0) &=& (x_1^0)^{s_1} \, (x_2^0)^{s_2} \,
\partial_{x_1^0}^{s_1} \, \partial_{x_2^0}^{s_2} 
\,,\nonumber\\[4mm]
T^P_s(x^0) &=& (x^0)^{s} \partial_{x^0}^{s}
\,.
\en


The results for the perturbative coefficients
parameterizing the NLP hadronic structure functions are
(we display only nonvanishing coefficients):

(1) for quark-antiquark annihilation 
\eq
R^{{\rm NLP}, L}_{qq;00} &=&  \rho^2 \, R^{{\rm NLP}, T}_{qq;00}
+ g_{q\bar q; 1} \ L_\rho = 2 \, g_{q\bar q; 1} \, (1 + L_\rho) \,,
\nonumber\\[5mm]
R^{{\rm NLP}, L}_{qq;01} &=& R^{{\rm NLP}, L}_{qq;10}
= \rho^2 \, R^{{\rm NLP}, T}_{qq;01} \,=\, \rho^2 \, R^{{\rm NLP}, T}_{qq;10}
= \rho \, R^{{\rm NLP}, \Delta}_{qq;01}
- g_{q\bar q; 1} \, (1 + L_\rho) 
\nonumber\\[2mm]
&=& - \rho \, R^{{\rm NLP}, \Delta}_{qq;10}
- g_{q\bar q; 1} \, (1 + L_\rho) 
= - g_{q\bar q; 1} \, (1 - L_\rho) \,, 
\nonumber\\[5mm]
R^{{\rm NLP}, L}_{qq;11} &=& 
\rho^2 \, R^{{\rm NLP}, T}_{qq;11}
= - 2 \, g_{q\bar q; 1} \, L_\rho \,,
\nonumber\\[5mm]
  P^{{\rm NLP}, L}_{qq,10}
&=& P^{{\rm NLP}, L}_{qq,20}
\,=\, \rho^2 \, P^{{\rm NLP}, T}_{qq,10} 
- g_{q\bar q; 1} \, \frac{1 + z^2}{2 (1-z)_{+}}
\,=\, - \rho^2 \, P^{{\rm NLP}, T}_{qq,20} 
- g_{q\bar q; 1} \, \frac{1 + z^2}{2 (1-z)_{+}}
\nonumber\\[2mm]
&=& \rho \, P^{{\rm NLP}, \Delta}_{qq,10}
- g_{q\bar q; 1} \, \frac{2 + z}{(1-z)^2_{+,1}}
\,=\, - \rho \, P^{{\rm NLP}, \Delta}_{qq,20}
- g_{q\bar q; 1} \, \frac{2 + z}{(1-z)^2_{+,1}}
\,=\, - g_{q\bar q; 1} \, \frac{2 + z + z^3}{2 (1-z)^2_{+,1}}
\,, 
\nonumber\\[5mm]
  P^{{\rm NLP}, L}_{qq,11}
&=& P^{{\rm NLP}, L}_{qq,21}
\,=\, \rho^2 \, P^{{\rm NLP}, T}_{qq,11} 
\,=\, \rho^2 \, P^{{\rm NLP}, T}_{qq,21}  
\nonumber\\[2mm]
&=&  \rho \, P^{{\rm NLP}, \Delta}_{qq,11}
+ g_{q\bar q; 1} \, \frac{1 + z}{(1-z)^2_{+,1}}
\,=\, - \rho \, P^{{\rm NLP}, \Delta}_{qq,21}
+ g_{q\bar q; 1} \, \frac{1 + z}{(1-z)^2_{+,1}}
\,=\, g_{q\bar q; 1} \, \frac{(1 + z) (1 + z^2)}{2 (1-z)^2_{+,1}}
\,. 
\en 

(2) For quark-gluon Compton scattering process:
\eq
R^{{\rm NLP}, L}_{qg;00} &=&
\rho^2 \, R^{{\rm NLP}, T}_{qg;00}
- 2 \, g_{q g; 1} \, (1 - 2 L_\rho) 
\,=\, \rho \, R^{{\rm NLP}, \Delta}_{qg;00} - g_{q g; 1} \,
(3 - 2 L_\rho) = - g_{q g; 1} \, \Big(\frac{5}{2} - 4 L_\rho\Big) 
\,,
\nonumber\\[5mm]
R^{{\rm NLP}, L}_{qg;10} &=&
\rho^2 \, R^{{\rm NLP}, T}_{qg;10} - 4 \, g_{q g; 1} \, L_\rho
\,=\, \rho \, R^{{\rm NLP}, \Delta}_{qg;10} -  6 \, g_{q g; 1} \, L_\rho
= - 5 \, g_{q g; 1} \,  L_\rho
\,,
\nonumber\\[5mm]
P^{{\rm NLP}, L}_{qg,10}
&=& \rho^2 \, P^{{\rm NLP}, T}_{qg,10} +
g_{q g; 1} \, \frac{4 z^2}{(1-z)^2_{+,1}}
\,=\, \rho \, P^{{\rm NLP}, \Delta}_{qg,10}
+ g_{q g; 1} \, \frac{2 z (2 + z)}{(1-z)^2_{+,1}}
= g_{q g; 1} \, \frac{z (1 + (1+z)^2)}{(1-z)^2_{+,1}} 
\,, 
\nonumber\\[5mm]
P^{{\rm NLP}, L}_{qg,20}
&=& \rho^2 \, P^{{\rm NLP}, T}_{qg,20} 
-  g_{q g; 1} \, \frac{1 + 5 z + 4 z^2 - 2 z^3}{2 (1-z)^2_{+,1}}
\,=\,
\rho \, P^{{\rm NLP}, \Delta}_{qg,20} -
g_{q g; 1} \, \frac{z (1 + z)}{(1-z)_{+}}
\,=\, -  g_{q g; 1} \,
\frac{(1 + z)^2}{(1-z)_+}
\,, 
\nonumber\\[5mm]
P^{{\rm NLP}, L}_{qg,21}
&=&
\rho^2 \, P^{{\rm NLP}, T}_{qg,21}  
\,=\, - \rho \, P^{{\rm NLP}, \Delta}_{qg,21} 
+ g_{q g; 1} \, \frac{1 + z}{(1-z)_{+}}
\,=\,  g_{q g; 1} \, \frac{(1 + z) (z^2 + (1 + z)^2)}{2 (1-z)_{+}}
\,.  
\en

\end{document}